\documentclass[range]{ar2e}%%%Put [range] option here for references citations to
\def\simgt{\lower 2pt \hbox{$\, \buildrel {\scriptstyle >}\over {\scriptstyle \sim}\,$}}
\def\simlt{\lower 2pt \hbox{$\, \buildrel {\scriptstyle <}\over {\scriptstyle \sim}\,$}}

\def\heao1{{\it {\it HEAO-1}\/}}

\begin{document}

\input epsf.tex 
\input epsf.def 
\input psfig.sty

\jname{Annual Reviews of Astronomy \& Astrophysics}
\jyear{2006}
\jvol{45}
\ARinfo{1056-8700/97/0610-00}

\title{Relativistic X-ray Lines from the Inner Accretion Disks Around Black Holes}

\markboth{Relativistic Disk Lines}{Relativistic Disk Lines}

\author{J. M. Miller$^1$
\affiliation{$^1$Department of Astronomy and Astrophysics, The
University of Michigan, 500 Church Street, Ann Arbor, Michigan, 48109,
USA; email: jonmm@umich.edu}}


\begin{abstract}
Relativistic \hbox{X-ray} emission lines from the inner accretion disk
around black holes are reviewed.  Recent observations with the {\it
Chandra X-ray Observatory}, {\it X-ray Multi-Mirror Mission-Newton},
and {\it Suzaku} are revealing these lines to be good probes of strong
gravitational effects.  A number of important observational and
theoretical developments are highlighted, including evidence of black
hole spin and effects such as gravitational light bending, the
detection of relativistic lines in stellar--mass black holes, and
evidence of orbital--timescale line flux variability.  In addition,
the robustness of the relativistic disk lines against absorption,
scattering, and continuum effects is discussed.  Finally, prospects
for improved measures of black hole spin and understanding the spin
history of supermassive black holes in the context of black
hole-galaxy co-evolution are presented.  The best data and most
rigorous results strongly suggest that relativistic X-ray disk lines
can drive future explorations of General Relativity and disk
physics.

\end{abstract}

\maketitle

%------------------------------------------------------------------------

\section{Introduction to Relativistic Disk Lines}

Accretion onto compact objects is the only viable means of powering
the sustained X-ray energy release observed in active galactic nuclei
(AGN) and in X-ray binaries.  At sufficiently high fractions of the
Eddington luminosity ($L_{X}/L_{Edd.} \geq 0.001$, and perhaps also at
smaller fractions), accretion onto both the supermassive black holes
in AGN and stellar-mass black holes is thought to proceed through a
standard Shakura--Sunyaev accretion disk (Shakura \& Sunyaev 1973).
As the inner disk can extend to within $6~GM/c^{2}$ of the black hole
even in the absence of black hole spin, it may serve as a probe of the
strong gravitational environment.  In AGN, the innermost accretion
disk spectrum peaks in the UV band; neutral hydrogen absorption can
make direct measurements of the disk continuum very difficult.  In
stellar-mass black holes, the inner disk emits in X-rays with a
typical temperature of $kT \simeq$1~keV.  In these cases, it is also
difficult to get a clean measure of disk properties, due to strong
Galactic line-of-sight absorption at low energy and simultaneous and
overlapping hard X-ray emission.  Moreover, in both AGN and
stellar-mass black holes, the observational reality is that
unambiguous signatures of gas orbiting close to a black hole -- such
as strong Doppler shifts and gravitational red-shifts -- are not
clearly revealed through continuum emission.

Owing only to a combination of abundance and fluorescence yield, Fe
K-shell emission lines are the strongest X-ray emission lines in most
AGN and X-ray binaries.  When the disk ionization is low, fluorescence
is the primary line emission mechanism; in highly ionized disks,
recombination dominates.  Inspired in part by the moderately broad
Fe~K line observed in the stellar-mass black hole Cygnus X-1 (Barr,
White, \& Page 1985), disk line models were calculated for two extreme
cases: a zero-spin Schwarzschild black hole (Fabian et al.\ 1989), and
a maximally-spinning Kerr black hole (Laor 1991).  Both models predict
a strongly asymmetric line profile with a red wing extending down to
low energy.  The Solid-state Imaging Spectrometer (SIS) aboard the
{\it Advanced Satellite for Cosmology and Astrophysics}
(or, {\it ASCA}) was the first X-ray spectrometer capable of revealing
multiple X-ray lines from abundant elements.  Moreover, the energy
resolution of the SIS ($E/dE \simeq 30$, approximately) was sufficient
to resolve the shape of X-ray lines.  Observations with {\it ASCA}
marked the true beginning of relativistic disk line studies.

In recent years, observations with {\it XMM-Newton} have led the way
in detecting and measuring the properties of relativistic disk lines
in Seyferts.  While {\it Chandra} offers the highest spatial and
spectral resolution and the lowest background, its low effective area
means that it is less suited for relativistic disk line studies.  The
high effective area of {\it XMM-Newton} is essential: not only is the
effective collecting area below the Fe K band higher than prior
imaging missions, but so too is the effective collecting area above
the Fe K band.  While the lower energy range is important for
detecting the red wing of a relativistic line, the energy range above
the line is essential for accurately measuring the continuum and disk
reflection spectra.  Moreover, accurately measuring the blue wing is
essential for constraining the inner disk inclination.  The high
effective area and broad-band coverage of {\it Suzaku} (0.5--700~keV)
position the observatory to make contributions to relativistic line
studies that will equal or exceed even those of {\it XMM-Newton}.

In the sections that follow, recent detections and applications of
relativistic accretion disk lines in the {\it Chandra}, {\it
XMM-Newton}, and {\it Suzaku} era are discussed in the context of
exploring general relativity and accretion onto black holes.  The need
to emphasize results which clearly demonstrate recent advances in the
field imposes strong limits; some relevant studies are only mentioned
in passing and others are omitted.  Indeed, accretion onto black
holes, accretion disks, the nature of spacetime close to black holes,
and probes of strong gravitational effects are all large topics and
deserving of more attention and explanation than is feasible herein.

Accretion onto black holes is treated thoroughly in Frank, King, \&
Raine (2002).  Excellent general treatments of active galactic nuclei
(AGN) are given by Peterson (1997) and Krolik (1999).  Remillard \&
McClintock (2006) present a useful review of the phenomenology
observed in stellar-mass black holes in X-ray binaries; this review
and others are more complete and representative with regard to black
hole spin diagnostics (see, e.g., Miller 2006).  Relativistic disk
lines are reviewed with specific attention to important theoretical
aspects in Reynolds \& Nowak (2003), and with attention to scattering
processes and atomic transitions in Liedahl \& Torres (2005).  The
reader is also referred to Fabian \& Miniutti (2006) for a discussion
of X-ray diagnostics of Kerr spacetime phenomena.

%......................................................................

\subsection{Initial Detections of Relativistic Disk Lines}

Seyfert-1 AGN are the class of supermassive black holes that offer the
cleanest view of the innermost accretion flow.  As such, Seyfert-1 AGN
and are the optimal set of targets for relativistic disk line studies.
Using the {\it ASCA}/SIS, Tanaka et al.\ (1995) first observed an
asymmetric disk line profile in the Seyfert-1 AGN MCG-6-30-15.  Other
detections followed thereafter, and were greatly aided by the
long--look observations made late in the {\it ASCA} mission.
Systematic analysis of {\it ASCA} spectra of AGN are detailed in
landmark survey papers by Nandra et al.\ (1997) and Reynolds (1997).
Importantly, these surveys suggested that disk lines are common in
Seyfert-1 AGN and should serve as a common diagnostic.

Photon counting rate limitations prevented the {\it ASCA}/SIS from
making clear detections of relativistic disk lines in stellar-mass
black holes.  Despite its low spectral resolution, the gas imaging
spectrometer (GIS) aboard {\it ASCA} could achieve detections of
relativistic disk lines, but an observational emphasis on SIS
spectroscopy served to ensure that few stellar-mass black holes were
observed at high mass accretion rates.  Indeed, although prior
observations with gas spectrometers suggested the presence of disk
lines in stellar-mass black holes, it was not until {\it Chandra}
observed Cygnus X-1 that a single dynamically--broadened line was
clearly revealed and separated from any narrow components (Miller et
al.\ 2002a).

%......................................................................

\subsection{Relativistic Lines and Disk Reflection: Theoretical Background}

In the case of white dwarfs, atmosphere models are a robust means of
measuring the stellar surface gravity (see, e.g., Liebert et al.\
2004).  In accreting black hole systems, it is a combination of
relativistic line and disk reflection models that is used to measure
gravitational and Doppler shifts, and thereby inner disk radii and
black hole spin.  Physically and practically, modeling white dwarf
atmospheres and black hole disk reflection are very similar.  In both
cases, the complex but entirely tractable details of gas processes and
scattering are treated in order to accurately reveal gravitational
effects.

A number of pure relativistic line models are available for spectral
fitting through packages like XSPEC and ISIS (Arnaud 1996, Houck \&
Denicola 2000).  The models are based on ray tracing: the Doppler
shifts and gravitational red-shifts imprinted on photons escaping from
points close to a black hole are built-up in libraries.  The most
important considerations are the spin of the black hole, the radius of
the inner edge of the accretion disk, the outer radius of line
emission, the inclination at which the inner disk is viewed by a
distant observer, and the line emissivity as a function of radius
($r^{-3}$ is expected based on geometric considerations, assuming a
standard thin accretion disk and isotropic source of disk
irradiation).

In principle, lines emitted from the inner disk may be used to measure
black hole spin because the innermost stable circular orbit (ISCO)
around a black hole depends on the spin of a black hole: $r_{ISCO} =
6.0~r_{g}$ for $a = 0$, and $r_{ISCO} = 1.25~r_{g}$ for a maximal
value of $a = 0.998$ (where $r_{g} = GM/c^{2}$ and $a = cJ/GM^{2}$;
see Bardeen, Press, \& Teukolsky 1972, and Thorne 1974).  The line
expected for an accretion disk orbiting a non-spinning Schwarzschild
black hole was described by Fabian et al.\ (1989); the maximal spin
Kerr case was described by Laor (1991).  It was not until 2004 that
models became available that included spin as a variable parameter
that could be constrained directly by spectra.  Such models have been
developed and implemented into XSPEC by Dovciak, Karas, \& Yaqoob
(2004), Beckwith \& Done (2004), and Brenneman \& Reynolds (2006).
The dependence of the ISCO on $a$ and representative line profiles are
shown in Figure 1.  In the future, numerical simulations of standard
thin accretion disks may be able to test analytic expectations for the
dependence of an effective ISCO on $a$ (present simulations can treat
hot thick disks; see Krolik \& Hawley 2002, Krolik, Hawley, \& Hirose
2005).

In practice, black hole spin measurements with relativistic lines also
depend on two principles that are upheld by observations and
physically sound arguments.  First, an accretion disk extends to the
ISCO for a given value of $a$ at all accretion rates above a certain
threshold.  Observations appear to strongly confirm this assumption.
Second, gas within the ISCO emits only weakly.  Theoretical arguments
suggest that the gas within the ISCO should be hot, nearly fully
ionized, and a negligible source of line emission (e.g. Young, Ross,
\& Fabian 1998; Brenneman \& Reynolds 2006).  Moreover, observed line
profiles are very different than those predicted if line emission from
within the ISCO is important (see, e.g., Reynolds \& Begelman 1997).

Iron emission lines are the most obvious reaction of an accretion disk
to irradiation by an external source of hard X-rays.  The second
important feature is an excess peaking between 20--30 keV due to
Compton back-scattering; when added to power-law emission, this
appears to be a flux excess that is sometimes called the ``reflection
bump'' (see the left panel in Figure 2).  An external X-ray source is
required because accretion disks do not efficiently self-irradiate;
their emission peaks are well below the energy required to remove K-shell
electrons from Fe atoms.  The overall process of disk illumination and
its response is known as disk reflection.  The most important
parameters in such models are the disk ionization state, the ratio of
incident to reflected emission (equivalent to the ``covering factor''
in some models), the temperature of the disk, and the inclination at
which the disk is viewed by a distant observer.

The reaction of a neutral disk to power-law irradiation is described
by George \& Fabian (1991).  (An especially useful point of reference
from this work is that reflection from a disk extending to the ISCO
should produce an Fe K line with an equivalent width of 180~eV.)  A
power-law is typically assumed in such models because it is a
reasonable representation of Comptonization, synchrotron, and/or
synchrotron self-Comptonization spectra.  Each of these processes
might contribute to hard X-ray emission in black holes, in varying
degrees.  Since this initial study, a number of generalizations and
refinements have followed (e.g. Ross, Fabian, \& Young 1999;
Nayakshin \& Kallman 2001).  These include the ability to handle high
ionization parameters, the ability to account for different sources of
hard X-ray irradiation, and different vertical density prescriptions
(see the right panel in Figure 2).  Most new reflection models
explicitly include line emission (some intermediate models excluded
the line components).  Reflection models are calculated in the fluid
frame, so they must be convolved with a relativistic line model in
order to match the spectrum received by a distant observer.  The
parameters derived from fitting such a convolved spectrum can be used
to constrain black hole spin; this procedure is merely a more complete
method of deriving fundamental parameters than fitting the Fe disk
line alone.

Many lines of evidence demand that hard X-ray emission in black hole
systems be central and compact.  The paradigm for many years was that
hard X-rays originate via inverse-Compton scattering in a diffuse, hot
corona.  Many disk lines imply a centrally-concentrated source of hard
X-ray emission; these findings and correlations between X-ray and
radio (jet) fluxes imply that processes in the base of a jet (direct
synchrotron and/or synchrotron self-Comptonization) may also be
important.  This suggests that the jet base may be only a few $r_{g}$
in size, but this is broadly consistent with recent models
(e.g. Markoff, Nowak, \& Wilms 2005).  However, the exact details do
not need to be known for disk reflection to be modeled effectively: as
long as a hard X-ray source illuminates the inner disk, reflection
results, and iron lines can be exploited to measure fundamental
quantities.  Details of emission mechanisms aside, in practice the
continuum X-ray emission in black holes is fairly simple.  In Seyfert
AGN, the continuum can be described with a simple power-law across the
X-ray bandpass.  In some cases, a ``soft excess'' is observed (see
below); this component does not strongly affect the Fe~K band.  In
stellar-mass black holes, both the disk and a non-thermal component
can contribute in the soft X-ray band.  However, these can be modeled
effectively, and no continuum emission apart from a disk and power-law
is statistically required in the strong majority of cases.  With the
addition of disk reflection and iron line components, stellar-mass
black hole spectra are well-described, and strong lines are robust
against different continuum models.  Representative Seyfert-1 and
stellar-mass black hole X-ray spectra are shown in Figure 3.

%----------------------------------------------------------------------

\section{Relativistic Disk Lines in Seyfert Actve Galactic Nuclei}

%------------------------------------------------------------------------

\subsection{Recent Studies of MCG-6-30-15}

The Seyfert-1 AGN MCG-6-30-15 is presently the single most important
source for studies of relativistic disk lines around black holes.
Spectra from MCG-6-30-15 provided the first clear example of
relativistic dynamics affecting the shape of a line (Tanaka et al.\
1995), and new observations with {\it Chandra}, {\it XMM-Newton}, and
{\it Suzaku} reveal strong evidence of black hole spin and possible
signatures of gravitational light-bending.  No other source has
received the attention and scrutiny that MCG-6-30-15 has received.
The strong line observed in its spectrum, and its relatively high
broad-band X-ray flux, have served to ensure that it is a favorite
target for observers.  The same factors also make it an ideal test
case for new disk line and reflection models.  MCG-6-30-15 is, and
will continue to be, the gold standard (however, devoting similar
attention to other sources is essential for continued progress in the
field).

The results of the first {\it XMM-Newton} observation of MCG-6-30-15
are presented in Wilms et al.\ (2001).  The observation occurred in
what may be called a ``deep minimum'' state (Iwasawa et al.\ 1996; see
also Reynolds et al.\ 2004), wherein the disk line and reflection
parameters are especially prominent relative to the continuum.  Fits
to the observed spectrum reveal an extremely skewed Fe disk line, with
a red wing extending down to $\sim3$~keV.  The line was fit well using
the Laor model with the inner radius extending down to $r_{in} =
1.23~r_{g}$ -- strongly suggestive of a very high black hole spin
parameter.  This finding confirms prior evidence of spin initially
revealed in especially long {\it ASCA} observations (Iwasawa et al.\
1996).  Importantly, fits to the iron line and reflection spectra in
this initial {\it XMM-Newton} observation clearly require a steep line
emissivity index ($q = 4.5$, recall $J(r) \propto r^{-q}$).  If the
hard X-ray source is an isotropic emitter, the steep emissivity
profile is consistent with magnetic energy extraction from a spinning
black hole dissipating extra energy in the inner accretion disk
(Blandford \& Znajek 1977; see also Agol \& Krolik 2000).

Long-look observations of MCG-6-30-15 have provided precise
measurements, and have proved to be important in establishing that the
broad emission features observed can only be plausibly explained as
emission lines shaped by relativistic effects (see Vaughan \& Fabian
2004, A. K. Turner et al.\ 2004, and the discussion below).  The
results of a 320~ksec observation of MCG-6-30-15 are reported in
Fabian et al.\ (2002).  Here again, fits imply an extremely high spin
parameter ($r_{in} = 2.0\pm 0.2~ r_{g}$) and enhanced inner energy
dissipation ($q = 4.8\pm 0.7$).  Fabian et al.\ (2002) included fits
with a broken power-law form for the line emissivity; at radii greater
than $6~r_{g}$, the emissivity is consistent with expectations
($q=3$).  Whereas the fits made by Wilms et al.\ (2001) were based
only on an {\it XMM-Newton} spectrum, the long-look observation
detailed by Fabian et al.\ (2002) was made simultaneously with {\it
BeppoSAX}, providing a continuum determination up to 100~keV.  The
broad-band spectrum also requires relativistically-skewed disk
reflection.  {\it Suzaku} observations of MCG-6-30-15 confirm results
based on {\it XMM-Newton} data (see Figure 4), and demonstrate that
explanations for line and continuum variability patterns that invoke
gravitational light bending extend to the broad-band disk reflection
continuum (Miniutti et al.\ 2007; more discussion of these results can
be found below).

MCG-6-30-15 is the only black hole wherein a direct spin
determination has been made.  This was achieved through fits with one
of the new variable--spin relativistic line profiles noted above.
Brenneman \& Reynolds (2006) report $a = 0.989^{+0.009}_{-0.002}$
(90\$ confidence errors) based on fits to {\it XMM-Newton} spectra.
This measurement is especially robust -- a conservative continuum
spectral model was used that allowed for the effects of absorption at
low energy (for more on the effect of low energy absorption, see the
sections below).  The result is of obvious importance by itself, but
it also demonstrates that fits with the maximal-spin Laor line model
that yield small inner radii do indeed serve as strong indications of
high spin (e.g. $a \geq 0.9$).  It is interesting to note that the
extremely high spin parameter suggested by these fits would rule out
spin-down through magnetic field coupling; that possibility was
suggested by Wilms et al.\ (2000) to account for the steep emissivity
parameter required by the line.

Some studies have concluded that relativistic disk lines cannot serve
as good indicators of black hole spin, because the nature of emission
within the plunging region is uncertain and may serve to give faulty
information on the ISCO (see, e.g., Dovciak et al.\ 2004).  This
conclusion appears to be largely incorrect.  Sensitive spectra may
indeed permit meaningful spin measurements.  Gas within the plunging
region is likely to be fully ionized, giving no line emission.
Indeed, Brenneman \& Reynolds (2006) constructed models forcing
emission from within the plunging region, and these models give
unphysical results when they are applied to the spectrum of
MCG-6-30-15.  It should also be noted that black hole spin is
essential to models that succeed in describing disk line and continuum
variability in black hole.  To the extent that variability can serve
as a check on spin measurements from direct line fitting, current
variability studies suggest that direct line fitting is yielding
accurate spin constraints (see, e.g., Miniutti \& Fabian 2004).

% ----------------------------------------------------------------------

\subsection{New Results from Well-Known Seyfert Active Galactic Nuclei}

{\it NGC~3516}: An analysis of initial {\it Chandra} and {\it
XMM-Newton} spectra is presented in Turner et al.\ (2002).  In
addition to the clear detection of relativistic disk line profile like
that found in MCG-6-30-15, the spectra appear to have {\it narrow} Fe
line components.  If the shifts of the narrow components is
interpreted in terms of Doppler shifts, enhanced emission from
$35~r_{g}$ and $175~r_{g}$ is implied; this may be consistent with
warps or with hard X-ray emission arising in distinct co-rotating
magnetic flares.  The detection of distinct narrow features in this
observation was the first observational realization of the potential
of Fe K lines to trace orbital-timescale variability in Seyfert AGN.
Subsequent studies have yielded important results that are discussed
in more detail below.  Fits to the relativistic disk line revealed in
{\it Suzaku} observations are detailed in Markowitz et al.\ (2006).

{\it NGC 4051}: Initial observations of NGC 4051 with {\it Chandra}
and {\it XMM-Newton} are reported in Uttley et al.\ (2003) and Uttley
et al.\ (2004).  Observations with {\it RXTE} were made simultaneously
with {\it Chandra}, and a good fit is obtained when a Laor line is
included in the full spectral model.  Indeed, fits with the Laor
component give $r_{in} = 1.33~r_{g}$, suggestive of a high black hole
spin parameter.  The {\it XMM-Newton} spectrum also shows evidence of
a relativistic disk line, but it is not modeled explicitly in Uttley
et al.\ (2004).  Ogle et al.\ (2004) explicitly fit the relativistic
line detected detected with {\it XMM-Newton} using a Laor model, and
measure $r_{in} \leq 2.1~r_{g}$.  This small inner radius is
consistent with the line results reported by Uttley et al.\ (2003),
and consistent with fits to a possible O~VIII line in the {\it
XMM-Newton} gratings spectrum (see below).  Spectral variability
analysis by Ponti et al.\ (2006) also suggests a very small inner disk
radius compatible with high spin.

{\it Markarian 766}: The first reports on {\it XMM-Newton}
observations of Mrk 766 focused on putative relativistic C, N, and O
disk lines (these features are discussed in a subsequent section).
Mason et al.\ (2003) report on fits to the relativistic Fe line in Mrk
766 using the Laor model; an inner disk radius of $r_{in} =
1.32^{+0.8}_{-0.1}$ is measured, suggesting that Mrk 766 may also
harbor a black hole with a high spin parameter.  More recently,
extremely deep observations of Mrk 766 with {\it XMM-Newton} have
revealed evidence for orbital-timescale variability in narrow Fe line
components (Turner et al.\ 2006).

{\it MCG-5-23-15}: {\it XMM-Newton} observations of the Seyfert 1.9
MCG-5-23-16 also reveal a relativistic disk line (Dewangan, Griffiths,
\& Schurch 2003; also see Balestra, Bianchi, \& Matt 2004).  It is
unusual to observe relativistic disk lines in Seyfert-2 AGN because
the torus prevents a clear view of the inner accretion disk.  In
MCG-5-23-16, the Fe~K line profile implies an inclination of
$47^{\circ}$ -- higher than that in sources like MCG-6-30-15 ($i \leq
30^{\circ}$).  Though the line profile is clear, its shape is not as
well determined as other relativistic iron lines, and there is no
statistical distinction between line models assuming zero spin and
maximal spin (Dewangan, Griffiths, \& Schurch 2003; Reeves et al.\
2007).  Figure 5 and Figure 6 depict the {\it Suzaku} spectrum of
MCG-5-23-16.

{\it NGC 5548}: In a few sources, indications for broad Fe K lines in
{\it ASCA} spectra have not been confirmed with new observatories, and
NGC 5548 is one example.  Independent systematic analyses
found evidence for broad Fe K emission lines in {\it ASCA} spectra of
NGC 5548 (Reynolds 1997, Nandra et al.\ 1997).  Pounds et al.\ (2003a)
analyzed an {\it XMM-Newton} observation of NGC~5548, and report no
evidence of a broad disk line (though an upper limit on such a
feature is not given).  The Fe K line emission is restricted to a
single narrow line from neutral iron, consistent with reflection from
the distant torus.  If the disk line is truly absent, it is difficult
to explain within the context of simple disk reflection geometries.
However, sources that have been studied more extensively -- such as
MCG-6-30-15 -- show strong variations in line strength and a
saturation of the line at high continuum flux levels.

{\it NGC 3783}: Spectra of NGC 3783 obtained by {\it ASCA} provided
some of the most compelling emission line profiles detected in that
era (see, e.g., Nandra et al.\ 1997 and Reynolds 1997).  Observations
with {\it Chandra} and {\it XMM-Newton} have revealed that the low
energy absorption in this Seyfert AGN is extremely complex, and may
have an influence even up to the Fe~K band (see Kaspi et al.\ 2002,
Reeves et al.\ 2004, and Yaqoob et al.\ 2005).  Fits to these new
spectra demonstrate that distinguishing the red wing of a relativistic
line from curvature due to low-energy absorption is difficult (but not
impossible) when low-energy absorption is strong.  Although NGC 3783
has a uniquely rich absorption spectrum, the realization that
low-energy absorption can plausibly affect the Fe K band fueled
attempts to explain relativistic disk lines in other sources as
absorption--induced modeling artifacts.  New individual observations
and survey work makes it clear that relativistic lines are robust and
required even when low-energy absorption is strong (Nandra et al.\
2006; Reeves et al.\ 2006).

%......................................................................

\subsection{New XMM-Newton and Suzaku Surveys}

Some initial analyses of Seyfert spectra obtained with {\it
XMM-Newton} were complicated by uncertainties in the instrumental
calibrations and background flux.  Other work over-estimated the role
of low-energy absorption in shaping the continuum in the Fe~K band
(again, see the discussion below).  With the instrument
characteristics and proper scope of low energy absorption better
understood, Nandra et al.\ (2006) have undertaken a systematic
analysis of Seyfert AGN spectra obtained with {\it XMM-Newton} (also
see Guainazzi, Bianchi, \& Dovciak 2006).  This analysis is both
rigorous and conservative: the spectral models are sophisticated and
are in no way biased to detect relativistic disk lines.  Apart from
reasonable continua, the spectral models include variable low energy
absorption, a narrow Gaussian emission line to account for distant
emission from the torus, relativistic disk line models, and disk
reflection.  An especially important result of this work is that broad
Fe lines consistent with a disk origin are detected in 73\% of the
sample of 30 objects; this is consistent with {\it ASCA} surveys
despite the more conservative approach, and clearly demonstrates that
disk lines are very common.

Nandra et al.\ (2006) measure relativistic disk lines that require
emission from within $20~r_{g}$ in nine Seyferts, including: NGC 2992,
MCG-5-23-16, NGC 3516, NGC 3783, NGC 4051, NGC 4151, Mrk 766, and
MCG-6-30-15.  In each of these cases, alternative explanations for the
red wing of the lines, including low energy absorption shaping the
continuum, are statistically rejected.  The results reported by Nandra
et al.\ (2006) differ slightly from others, in that only marginal
evidence for black hole spin is found (in NGC 3783 and NGC 4151).
This is likely because a specific prescription for the disk
emissivity is enforced: $q=0$ within a break radius, and $q=3$ at
greater radii.  This choice does not affect the detection of
relativistic disk lines, but it does serve to affect constraints on
the inner disk radius and therefore black hole spin.  Analyses that
allow the data to determine the emissivity do find strong evidence of
spin (e.g. Fabian et al.\ 2002, Brenneman \& Reynolds 2006).  As
the field is still developing a picture of the corona and how it might
illuminate the disk, allowing the data to decide is the best possible
approach.

New observations of Seyfert AGN with {\it Suzaku} have also clearly
revealed relativistic disk lines in a number of sources.  {\it Suzaku}
is uniquely able to cover an energy range extending an order of
magnitude below and above the Fe~K band.  This means that the Compton
back-scattering hump peaking between 20--30 keV cannot merely be
detected, but rather it can be measured well.  Detections of the
reflection hump demand that the correct description of the spectrum
must include a disk line and reflection.  In an initial survey of
seven Seyfert AGN, Reeves et al.\ (2006) report the detection of
relativistic Fe disk lines and reflection in six sources: MCG-5-23-16,
MCG-6-30-15, NGC 4051, NGC 3516, 3C 120, and NGC 2992.  In one
case, NGC 2110, there is no evidence of a relativistic disk line
or reflection.  This {\it Suzaku} detection fraction (86\%) is thus
broadly consistent with the {\it XMM-Newton} results reported by
Nandra et al.\ (2006).

A few {\it Suzaku} results are particularly worthy of note.  Spectra
of NGC 3516 clearly reveal a relativistic disk line (the line is
required at more than 99.999\% confidence), despite relatively complex
low-energy absorption and narrow emission components within the Fe~K
band (Markowitz et al.\ 2006; Reeves et al.\ 2006).  Spectra of
MCG-6-30-15 reveal relativistically-skewed disk reflection, consistent
with the relativistic iron line detected in {\it XMM-Newton}
observations (Miniutti et al.\ 2007).  Moreover, the {\it Suzaku}
observations show that the disk reflection spectrum follows the same
variability pattern as the iron line; not only does this tie these
features together in the way anticipated theoretically, but it is
additional evidence that gravitational light bending may drive much of
the observed flux variability (see below).  The same variability
pattern is observed in {\it Suzaku} observations of NGC 4051 and
MCG-5-23-16, further supporting this picture (Reeves et al.\ 2006,
2007; see also Ponti et al.\ 2006).

% ......................................................................

\subsection{Individual Carbon, Nitrogen, and Oxygen Lines}

As noted above, iron is merely the most prominent disk reflection line
owing to a combination of its abundance, fluorescent yield, and
ability to retain electrons at high temperature.  For a plausible
range of ionization parameters (up to $\xi = 10^{3}$, where $\xi =
L/nr^{2}$; see, e.g., Ballantyne, Ross, \& Fabian 2002), disk
reflection models predict that relativistic O VIII disk lines should
be visible as weak (30--70~eV in equivalent width, assuming solar
metallicity) spectral features on top of the continuum.  Detecting
such lines in many Seyfert AGN is complicated by the presence of a
warm absorber in soft X-rays.  A fair assessment of present results is
that there is evidence in favor of such lines, but no conclusive
detections.

A major finding of the {\it ASCA} era was that many Seyfert-1 AGN have
low-energy spectral complexity consistent with ionized absorption
(Reynolds 1997).  The spectra of these "warm absorbers" strongly
suggests that O VII and O VIII contribute significantly to the total
opacity.  Higher resolution observations with the {\it Chandra}/HETGS
and {\it XMM-Newton}/RGS have since confirmed this in many cases, and
strongly suggest that warm absorbers are related to disk winds.  In an
analysis of the first {\it XMM-Newton}/RGS spectra of MCG-6-30-15 and
Markarian 766, however, Branduardi-Raymont et al.\ (2001) proposed
that the apparent edges seen in the spectra are actually the blue wing
of relativistic C VI, N VII, and O VIII lines.  This interpretation
partially hinged on the finding that the O VII edge was either absent
or red-shifted by 16,000~km/s; the latter was deemed unlikely as no O
VII absorption lines were found to be similarly red-shifted.  This
emission line model does fit the strongest features in the soft
X-ray bandpass well, and the small inner disk radii inferred ($r =
1.24~r_{g}$) are consistent with Fe~K line results (Wilms et al.\
2001; Fabian et al.\ 2002).  However, this model alone does not
account for a number of observed absorption lines.

An analysis of the initial {\it Chandra}/HETGS spectrum of MCG-6-30-15
found the O VII edge at its rest wavelength, and attributed the
apparent red-shift of the edge in the {\it XMM-Newton} spectrum to a
combination of O VII lines and Fe L absorption (Lee et al.\ 2001).
The model adopted to describe the {\it Chandra} spectrum requires that
some of the iron be embedded in dust within a highly ionized
medium in order to match optical reddening.  This "dusty warm
absorber" model produced a good fit to the {\it Chandra}/HETGS
spectrum without requiring strong relativistic C VI, N VII, or O VIII
lines.

Although both initial spectra were of limited sensitivity and both
models are physically motivated, the strong discrepancy between these
interpretations sparked a vigorous debate.  In fact, a combination of
both models may give the best description of the data.  A thorough
re-analysis of the {\it XMM-Newton}/RGS spectra of MCG-6-30-15 and
Markarian 766 by Sako et al.\ (2003) finds that relativistic lines
still dominate, but that a combination of relativistic lines and a
dusty warm absorber is required.  The relativistic O VIII disk line
measured by Sako et al.\ (2003) has an equivalent width of
$162\pm8$~eV; this is higher than predicted by disk reflection models
(e.g., Ballantyne, Ross, \& Fabian 2002).  Ogle et al.\
(2004) report evidence for a relativistic O VIII line in {\it
XMM-Newton} spectra of NGC 4051 with an equivalent width of
$\sim$90~eV, which is more consistent with current models.

At present, the definitive statement on the possibility of
relativistic lines from elements other than iron likely comes from an
analysis of the 320~ksec {\it XMM-Newton} long--look observation of
MCG-6-30-15.  A. K. Turner et al.\ (2004) find that the best-fit dusty warm
absorber model leaves residuals consistent with relativistic lines,
that the best-fit relativistic lines model requires dusty warm
absorption, and that neither model (by itself) is a statistically
superior description of the time-averaged spectrum.  In a procedure
borrowed from analysis of continuum versus iron K line variability,
Turner et al.\ (2004) derive a high-flux minus low-flux difference
spectrum to illustrate that absorption must be the dominant effect at
low energy, and that relativistic lines (O VIII in particular) are
constrained to have equivalent widths of approximately 30~eV or less
(broadly consistent with theoretical expectations).

In summary, relativistic disk lines from elements such as carbon,
nitrogen, and oxygen are predicted by disk reflection models, and it
should be possible to detect them even though they are weak.  By
virtue of its (relatively high) energy, O~VIII should be the most
easily detected soft X-ray line within a typical X-ray CCD or X-ray
calorimeter bandpass.  The longest observations and most careful
analyses provide some initial evidence for relativistic O VIII disk
lines in cases where relativistic Fe K lines are clearly detected.
Evidence for other lines is considerably weaker, especially after
accounting for the effects of warm absorbers, which appear to be the
dominant source of spectral complexity at low energy in Seyfert-1 AGN.
Clearly detecting relativistic O VIII lines is just within the reach
of {\it XMM-Newton} and {\it Suzaku}, and should be easily possible
with planned missions such as {\it Constellation-X} and {\it XEUS}.
Such lines will offer an important check on black hole spin
measurements and results related to strong field effects such as
gravitational light bending.

%------------------------------------------------------------------------

\subsection{The Soft Excess in Seyfert-1 AGN}

It has recently been realized that the ``soft excess'' flux observed
in addition to a simple power-law continuum in some Seyfert-1 AGN
(Arnaud et al.\ 1985) is not thermal flux arising from the accretion disk.
Such soft flux components are especially common in ``narrow-line''
Seyfert-1 AGN, which are thought to harbor lower mass black holes than
standard Seyfert-1 AGN.  Not only are the temperatures inferred from
blackbody fits to these components unreasonably high ($kT
\simeq$0.1--0.2~keV), but the temperatures do not vary with luminosity
(e.g., Gierlinksi \& Done 2004).  This has recently led to attempts to
describe the soft X-ray excess in terms of discrete atomic features.

The energetics and timescales observed from accretion onto compact
objects demand that X-ray emission is extremely
centrally--concentrated.  As a result, matter at hundreds or thousands
of Schwarzshild radii will see the central engine as a point source.
(If the X-ray emission is strongly scattered in a large volume around
the black hole, the point source approximation may be invalid;
however, many spectroscopic and timing studies suggest that this is
unlikely.)  Therefore, X-ray absorption is only sensitive to
velocities directly along the line of sight connecting the central
engine and the observer.  Accreting matter tends to move perpendicular
to this line of sight, unless it is being expelled in a wind.  It is
unlikely that the soft excess can be due to blurred absorption (e.g.,
Gierlinski \& Done 2004), then, as encoding tangential velocities in
absorption is extremely difficult, and a broad range of high
velocities both toward and away from the observer along a single line
of sight would be required to blur away individual absorption
features.

Instead, the soft X-ray excess in some Seyfert-1 AGN may be due to a
collection of blurred soft X-ray emission lines from the accretion
disk (e.g., Ballantyne, Iwasawa, \& Fabian 2001; Crummy et al.\ 2006).
This requires a low ionization parameter so that many disk lines can
be produced and blended together by relativistic blurring.  Fits to a
sample of Seyfert-1 AGN with a blurred disk reflection model for the
soft excess suggest that near-maximal spin is preferred in the strong
majority of cases (Crummy et al.\ 2006).  This is broadly consistent
with expectations (Volonteri et al.\ 2005), and may serve as a check
on results based on Fe K lines.  However, clear detections of more
distinct low--energy lines (from more ionized disks) in the future
would likely provide a favored complement to relativistic iron line
results, as individual line features bear the hallmarks of
relativistic shifts more clearly than blends which approximate a
continuum.

%------------------------------------------------------------------------

\section{Relativistic Lines from Disks Around Stellar--Mass Black Holes}

\subsection{Lines Detected Prior to Chandra and XMM-Newton}

As noted previously, it was a broad Fe~K emission line in the spectrum
of the stellar-mass black hole X-ray binary Cygnus X-1 (Barr, White,
\& Page 1985) that incited much of the original theoretical work on
relativistic disk lines and the subsequent observational studies of
Seyfert galaxies.  Until recently, however, studies of disk lines in
stellar-mass black holes progressed more slowly than in the case of
Seyfert AGN.  An {\it EXOSAT} spectrum of 4U 1543$-$475 revealed a
broad, asymmetric iron line (van der Woerd, White, \& Kahn 1989; see
also Park et al.\ 2004).  Done \& Zycki (1999) looked at many spectra
of Cygnus X-1 and reported evidence of relativistic smearing in the
broad-band reflection spectrum, not merely the Fe~K emission line.
Similar results were reported in V404 Cyg (Zycki, Done, \& Smith
1999).  Based on {\it RXTE} data, Balucinka-Church and Church (2000)
reported the detection of an extremely broad Fe~K emission line in
spectra of the stellar-mass black hole GRO~J1655$-$40, and Miller et
al.\ (2001) reported a broad Fe~K emission line in spectra of
XTE~J1748$-$288.

Although each of these results made progress, they were achieved using
gas detectors with characteristic energy resolutions of $\sim$5
($E/dE$, or $0.2c$).  Therefore, these results were less compelling
than the detections and asymmetries seen in {\it ASCA} CCD spectra of
Fe~K lines in Seyfert AGN.  Efforts to study the Fe~K line in Cygnus
X-1 using the {\it ASCA}/SIS were severely hampered by photon pile-up
(pile-up is the result of two photons registering as one because they
arrive within a single CCD frame time; see, e.g., Ebisawa et al.\ 1996).

The best studies of relativistic disk lines in stellar-mass black
holes prior to studies with {\it Chandra} and {\it XMM-Newton}
targeted GRS~1915$+$105 (Martocchia et al.\ 2002).  Broad-band spectra
obtained with {\it BeppoSAX} display prominent and skewed Fe~K disk
lines.  Fits to these lines imply a truncated accretion disk, thus
offering no evidence of black hole spin.  (These results are
consistent with later observations made using {\it XMM-Newton}; see
Martocchia et al.\ 2006.)  An extremely high inner disk ionization
may inhibit the detection of spin in GRS~1915$+$105.  Taken at face
value, however, these results call into question subsequent claims for
maximal spin based on disk continuum fits (McClintock et al.\ 2006).
It is worth noting again that disk continuum fits cannot unambiguously
reveal gravitational shifts and Doppler shifts, and so must be
regarded with caution.

\subsection{The Chandra and XMM-Newton Era}

{\it Chandra} and {\it XMM-Newton} have CCD spectrometers with fast
clocking modes suited to observing sources with high fluxes, free of
photon pile-up.  This instrumental advance made it possible to clearly
resolve broad iron lines in stellar-mass black holes, and to know
that broad features seen at low resolution are indeed relativistic
lines (rather than, e.g., a collection of narrow lines).  A {\it
Chandra}/HETGS observation of Cygnus X-1 clearly revealed a
relativistic line in a stellar-mass black hole for the first time
(Miller et al.\ 2002a).  The line profile in Cygnus X-1 can be fit
acceptably with a zero-spin line model (this is confirmed with {\it
XMM-Newton}; see also Miller 2006), suggesting that this system may
harbor a black hole with low or modest spin.  This is broadly
consistent with the expectation that accretion should drive-up spin;
Cygnus X-1 is a young binary with an O9.7 Iab supergiant donor star.

An observation of XTE~J1650$-$500 with {\it XMM-Newton} revealed an
Fe~K line that is extremely broad, and likely requires a very high
spin parameter (Miller et al.\ 2002b).  The line is well described by
the maximal-spin Laor line profile, and fits require $r_{in} \leq 2
r_{g}$.  The line emissivity is found to be very steep, with $q \simeq
5$.  Both of these parameters are similar to those measured in {\it
XMM-Newton} and {\it Suzaku} observations of the Seyfert-1 AGN
MCG-6-30-15 (Wilms et al.\ 2001; Fabian et al.\ 2002; Miniutti et al.\
2007).  The {\it XMM-Newton} measurements are confirmed and extended
in an analysis of three {\it BeppoSAX} spectra of XTE~J1650$-$500
(Miniutti, Fabian, \& Miller 2004).  Focusing only on the 2--10~keV
portion of the {\it BeppoSAX} spectra, Done \& Gierlinki (2006) note
that a highly ionized wind could reduce the breadth of the line;
however, this result is based on the low resolution spectra and a
narrow bandpass, and is inconsistent with both the {\it XMM-Newton}
results and the results of variability studies (Miniutti, Fabian, \&
Miller 2004; Rossi et al.\ 2005). 

Tbe recurrent transient GX~339$-$4 is to relativistic disk line
studies in stellar-mass black holes as MCG-6-30-15 is to similar
studies in Seyfert AGN.  It has been observed on multiple occasions,
at different fluxes, in different states, and with both {\it Chandra}
and {\it XMM-Newton}.  In all cases where the hard X-ray flux is
prominent, a strong relativistic Fe~K disk line is detected that is
consistent with a high degree of black hole spin.  Each relevant
spectrum of GX~339$-$4 has been fit with both phenomenological
continuum plus relativistic line models and more physically-motivated
disk line plus reflection models.  In each case, simultaneous
observations with {\it RXTE} have been employed to define the
broad-band continuum and overall disk reflection spectrum.
 
{\it Chandra} first observed GX~339$-$4 in outburst, while the source
was in an ``intermediate'' state.  Fits with the Laor relativistic
line plus ionized disk reflection model give $r_{in} =
1.3^{+1.7}_{-0.1}~ r_{g}$ (Miller et al.\ 2004b).  Similar fits to a
75~ksec observation of GX~339$-$4 in the ``very high'' and
``low/hard'' states give $r_{in} = 2-3~ r_{g}$ and $r_{in} =
3.0-5.0~r_{g}$, respectively (Miller et al.\ 2004a, 2006a).  The
``very high'' state spectrum is of particular interest, because the
spectral and variability phenomena observed in this state most closely
resemble the behavior seen in Seyfert AGN (please see Figure 7).
Similar to the case of XTE~J1650$-$500, the relativistic disk line
seen in the very high state of GX~339$-$4 requires
centrally--concentrated energy dissipation ($q = 5.5^{+0.5}_{-0.7}$).

An extremely skewed feature was recently reported in {\it XMM-Newton}
spectra of GRO~J1655$-$40, in which X-ray variability studies also
find evidence for spin (see, e.g., Strohmayer 2001, Torok et al.\
2005).  In two observations, Diaz Trigo et al.\ (2006) report evidence
for a relativistic line with $r_{in} = 1.4-1.5~ r_{g}$.  During the
time that {\it XMM-Newton} made these observations, simultaneous
observations with {\it RXTE} were made that are difficult to fit with
standard broad-band reflection models.  It is possible that different
continuum modeling would serve to reduce the extremity of the measured
line parameters.

The fact that relativistic disk lines are clearly revealed at moderate
and high resolution gives confidence in prior results obtained with
gas detectors.  Many analyses using gas spectrometer data were soon to
follow after the relativistic line in Cygnus X-1 was confirmed.
Observations with {\it BeppoSAX} revealed broad iron lines in the
black hole candidates 4U 1908$+$094 (in 't Zand et al.\ 2002) and
SAX~J1711.6$-$3808 (in 't Zand et al.\ 2002b).  A review of archival
{\it ASCA}/GIS data revealed relativistic disk lines in Cygnus X-1,
GRO~J1655$-$40, and GRS~1915$+$105, confirming prior results, and
discovered a disk line in XTE J1550--564 (Miller et al.\ 2005).
Please see Figure 8 for examples of lines found in archival data.
Physically-motivated fits to these spectra with the Laor line model
and ``pexriv'' reflection model (Magdziarz \& Zdziarski 1995) measure
$r_{in} = 1.6^{+4.0}_{-0.4}$ and $r_{in} = 1.8^{+0.5}_{-0.6}$ in
XTE~J1550$-$564 and GRO~J1655$-$40, respectively.  This suggests that
XTE~J1550$-$564 may harbor a rapidly spinning black hole, and
strengthens existing evidence for high spin in GRO~J1655$-$40.  As
with prior fits, there is no strong evidence for spin in
GRS~1915$+$105 based on its relativistic disk line.

% ----------------------------------------------------------------------

\section{Line Variability -- A New Frontier}

% ----------------------------------------------------------------------

\subsection{Relativistic Iron Line Flux Variations}

Multi-wavelength flux variability is seen in all accreting sources,
and the fastest X-ray variability from black holes is expected to
arise closest to the black hole itself.  Because relativistic disk lines
arise through illumination of the disk in this region, it is expected
that variations in the relativistic iron line and reflection spectrum
should always follow variations in the hard X-ray continuum in direct
proportion.  That is, relativistic iron line variations should follow
variations in the continuum, and should show the same degree of
variability.  These expectations are most easily tested in Seyfert-1
AGN.  In low flux phases, this expectation is borne out by
observations.  Yet, in higher flux phases, relativistic iron lines
appear to be less variable than the continuum.  The influence of
gravitational light bending close to a spinning black hole provides an
exciting explanation of this apparent discrepancy.  Studies of line
variability in stellar-mass black holes also affirm that broad iron
disk lines arise very close to the black hole.

Sensitive new data and clever analysis techniques reveal that in low
continuum flux states, broad iron lines vary with the continuum flux
in the manner predicted by simple reflection models.  Vaughan \&
Fabian (2004) and Ponti et al.\ (2004) employed {\it XMM-Newton}
observations and rms variability spectra to show that the red wing of
the iron line is more variable than the local continuum when
MCG-6-30-15 is in low flux phases (see also Reynolds et al.\ 2004).
Indeed, these findings also serve to confirm the expectation that the
narrow core seen in some relativistic line profiles either originates
in the outer disk or torus (for a recent discussion, see Nandra 2006);
the narrow line core is less variable than the local continuum.

Although the red wing of iron disk lines is more variable in low flux
phases, rms variability spectra show the opposite trend in periods of
relatively high flux.  Indeed, although the relativistic line profile
is clearly variable in time-selected spectra, the variability appears
to be driven more by variations in the power-law continuum than by the
line itself (e.g., Vaughan \& Fabian 2004).  In a prior analysis of a
long {\it ASCA} observation of MCG-6-30-15, Shih, Iwaswa, \& Fabian
(2002) had also found that the relativistic iron line appears to be
less variable than the continuum.  As noted above, appeals to complex
and/or variable absorption models cannot reconcile the apparently
disjoint variability behaviors in low and high flux phases:
flux-selected spectra show no evidence of progressive partial covering
(Vaughan \& Fabian 2004), and 47 of 51 absorption lines seen with {\it
XMM-Newton} do not vary between high and low flux states (A. K. Turner et
al.\ 2004).  Evidence that absorption is not driving the variability
is bolstered by new results from {\it Suzaku} observations of
MCG-6-30-15 which reveal that the entire disk reflection spectrum up
to 45~keV (where photoelectric absorption is unimportant) is less
variable than the continuum (Miniutti et al.\ 2007).

Apparently disjoint disk line and continuum flux correlations can be
explained with natural modifications that reflect the emerging
prevalence of black hole spin and evolving ideas concerning the
nature of the hard X-ray corona.  The extended red wings in some
relativistic disk lines signal black hole spin, and the effects of
gravitational light bending are especially strong near to spinning
black holes.  Light bending is largely ignored in simple disk reflection
models, but it is important because light bending can strongly affect
the flux that the disk and a distant observer will receive from a
constant-luminosity source (a magnetic flare above the disk, the base
of a jet, etc.).  Flux variability is often assumed to be due to a
source with varying intrinsic luminosity but static size and distance
from the black hole -- but this is only assumption.

New studies show that gravitational light bending can account for the
changing disk line and continuum flux patterns observed in a growing
number of black holes, including: MCG-6-30-15 (Miniutti et al.\ 2003;
Miniutti \& Fabian 2004), NGC 4051 (Ponti et al.\ 2006; see also
Reeves et al.\ 2006), 1H 0707-495 (Fabian et al.\ 2004), 1H 0419$-$577
(Fabian et al.\ 2005), MCG-5-23-16 (Reeves et al.\ 2006), and in the
stellar-mass black hole XTE J1650$-$500 (Miniutti, Fabian, \& Miller
2004; Rossi et al.\ 2005).  More examples are likely to be revealed
through analysis of new and existing {\it XMM-Newton} and {\it Suzaku}
observations.  Within the context of the light bending model, line and
continuum fluxes are positively correlated when the illuminating X-ray
source is within $\simeq 4~r_{g}$ of the black hole, weakly correlated
or uncorrelated when the illuminating source is within 4--13~$r_{g}$,
and uncorrelated or anti-correlated for greater source heights.  A
centrally-concentrated corona, consistent with the base of a jet or
corotating magnetic flares above the disk, is a secondary but
important implication of this model (flares and the base of a jet may
be the same thing, of course).  The success of this model is extremely
suggestive, not only of the importance of light bending, but of the
nature of the corona as well.  These results require a
centrally--concentrated corona, consistent with the base of a jet or
co-rotating magnetic flares above the disk (flares and the base of a
jet may be aspects of the same thing, of course).  Figure 9
illustrates the effects of gravitational light bending, and Figure 10
shows evidence for light bending in NGC 4051 and XTE J1650$-$500.

It does not appear likely that more careful treatment of e.g. the disk
ionization state would explain these observations without invoking
light bending, but more work is needed.  More examples of variability
consistent with light bending are likely to be revealed through
analysis of new and existing {\it XMM-Newton} and {\it Suzaku}
observations.  Importantly, the proximity of hard X-ray emission close
to the black hole implied by these and other results suggests that
reverberation mapping with {\it Constellation-X} and/or {\it XEUS} is
likely to be a highly successful probe of General Relativity.

%----------------------------------------------------------------------

\subsection{Orbital-timescale Line Variations in Seyfert Active Galactic Nuclei}

The next generation of major X-ray missions ({\it Constellation-X,
XEUS}) will have the collecting area and energy resolution required to
reveal the ISCO around supermassive black holes in unprecedented
detail using Fe~K line variability studies.  This will enable
scientists to better explore the predictions of General Relativity
near to black holes, in part via reverberation mapping (see below)
studies.  The {\it XMM-Newton} and {\it Suzaku} X-ray observatories
have the collecting area and sensitivity needed to provide early
indications of orbital-timescale variability and evidence for orbiting
"hot-spots" or disk turbulence, as shown in recent observations.

The first evidence for narrow and variable Doppler-shifted lines
in the X-ray spectra of NGC 3516 and Mrk 766 was found by Turner et
al.\ (2002) and Turner, Reeves, \& Kraemer (2004).  These narrow line
components are consistent with the transient shifts expected when a
disk reacts to an illuminating flare, or may indicate that certain
radii close to the black hole are illuminated preferentially.  Other
examples of such lines have recently been detected with {\it
XMM-Newton} and {\it Chandra} (see, e.g., Guainazzi 2003; Yaqoob et
al.\ 2003; Dovciak et al.\ 2004; Iwasawa et al.\ 2004).  It should be
noted that in all cases, the narrow red-shifted lines are relatively
weak, and do not contribute significantly to a relativistic line
profile.

The most compelling case for orbital-timescale variability in narrow
red-shifted Fe~K lines is found in an analysis of {\it XMM-Newton}
spectra of NGC 3516 presented by Iwasawa, Miniutti, \& Fabian (2004).
After the flux of the (relatively stable) relativistic emission line
is modeled and subtracted, nearly four 25~ksec cycles of a line
component moving between 5.7~keV and 6.5~keV are observed.  The
variability pattern revealed in a flux excess versus time map has a
distinctive saw-tooth pattern, which is consistent with Doppler shifts
and gravitational red-shifts acting on orbits close to the black hole.
The cycles are consistent with emission between 7--16~$r_{g}$, either
due to a co-rotating (magnetic) flare or turbulence in the accretion
disk (Iwasawa, Miniutti, \& Fabian 2004; Armitage \& Reynolds 2003).
It should be noted that the observed variability implies a black hole
mass of (1--5)$\times 10^{7}~M_{\odot}$, in agreement with
reverberation mapping results (Onken et al.\ 2003).  The observed
variations are found to be significant at the 97\% level of confidence
through extensive Monte Carlo simulations; this estimate is
conservative because the probability of cyclic variability is very low.
The cyclic variations detected in NGC 3516 are shown in Figure 11.

Cyclic variations in narrow Fe~K lines may also be present in {\it
XMM-Newton} observations of Markarian 766 (Turner et al.\ 2006).  The
flux excess versus time map is considerably more complex and noisy
than that found in NGC 3516.  The strongest variation seen corresponds
to a period of 165~ksec, or a radius of approximately 115~$r_{g}$ for
a black hole mass of $4.3\times 10^{6}~M_{\odot}$ (Turner et al.\
2006; Wang \& Lu 2001).  This is partially at odds with a growing body
of evidence (including relativistic line studies) that disk irradiation
in black holes is strongly concentrated in the very innermost regions.
The evidence for orbital variations in Markarian 766 is less
compelling than that in NGC 3516, but serves to illustrate the
potential power of such investigations.

%----------------------------------------------------------------------

\subsection{Orbital-timescale Line Variations in Stellar-Mass Black Holes}

Quasi-periodic oscillations (QPOs) are common features in the power
spectra of accreting stellar-mass black holes and neutron stars (for a
review, see van der Klis 2006).  So-called "kHz" QPOs in millisecond
X-ray pulsars are closely tied to the neutron star pulsation (spin)
frequency, and the highest frequency oscillations in stellar-mass
black holes (e.g., 100--450~Hz) are commensurate with Keplerian
orbital frequencies at the ISCO.  These facts signal that QPOs are
very likely disk frequencies, regardless of whether a given QPO
represents an orbital, precession, or resonance frequency.  Lower
frequency QPOs (somewhat arbitrarily, 10~Hz and below) are more common
and can be much stronger than high frequency QPOs.  If they are
Keplerian frequencies, slower QPOs relate to radii as large as a few
hundred $r_{g}$; however, the energy dependence of these X-ray QPOs
argues strongly that they arise closer to the compact object and thus
represent super-orbital frequencies.

In some low-flux states, GRS~1915$+$105 displays exceptionally strong
(10\% rms, and higher) and stable low-frequency QPOs.  Miller and
Homan (2005) exploited the stability of strong 1~Hz and 2~Hz QPOs
(found in separate observations) to extract phase-selected spectra,
and find that the broad Fe~K emission line in GRS~1915$+$105 varies
with QPO phase.  This result employed both direct spectral fitting and
difference spectra, the latter demonstrating the line variability in a
model-independent way.  If these QPOs only represent Keplerian
frequencies, the phase-dependence of the lines signals that they
likely originate from within $170\pm 50~r_{g}$ and $100\pm 30~r_{g}$
(respectively).  At these radii, relativistic shaping of lines is {\it
inevitable}.  The phase dependence observed is consistent with a warp
or precessing ring in the inner disk, and may even be a signature of
Lense-Thirring precession (see Markovic \& Lamb 1998; Fragile, Miller,
\& Vandernoot 2005; Schnittman, Homan, \& Miller 2006).  This result
builds on prior work which clearly demonstrated that Fe~K lines are
sensitive to variability frequencies (see, e.g., Gilfanov, Churazov,
\& Revnivtsev 2000); though prior efforts did not use QPOs or averaged
over phase.

The result linking Fe~K lines and QPOs is notable for a few reasons.
First, it shows that Fe~K lines must originate in the inner disk and
be shaped by relativistic effects in a model-independent way.  Second,
this connection provides a unique way of mapping the inner accretion
disk around stellar-mass black holes and neutron stars.  Third -- and
perhaps most importantly -- this result is analogous to detections of
orbital-timescale line variability in Seyfert-1 AGN such as NGC 3516
(Iwasawa, Miniutti, \& Fabian 2004) in that it is becoming possible to
explore inner disk structures, not just relativity, with Fe~K emission
lines.  Future X-ray missions such as {\it Constellation-X} and {\it
XEUS} will be able to exploit the Fe~K--QPO connection to map inner
disk structures and explore relativistic effects in a much greater
number of systems.

%......................................................................

\section{Alternatives to Relativistic Lines and Dynamical Broadening}

\subsection{The Effects of Warm-Absorber Disk Winds on Relativistic Lines}

As noted previously, whereas O VII and O VIII edges were sufficient to
describe most of the "warm absorbers" detected in some Seyfert-1 AGN
with {\it ASCA} (Reynolds 1997), the dispersive high-resolution
spectrometers aboard {\it Chandra} and {\it XMM-Newton} revealed
complex spectra rich in absorption lines and edges.  This realization
led to some charged claims that the red wing of relativistic Fe K
lines might generally be the result of poor fits to curvature induced
by low-energy absorption.  The initial claims were largely restricted
to conferences and were never published as actual fits to data in a
journal paper (they are partially summarized a Ph.D. thesis; see
Kinkhabwala 2003).  However, a few scientists later explored this
possibility in formal papers.  After careful analysis of deep
observations, and with added continuum constraints from {\it Suzaku},
it is clear that curvature in the Fe~K band is properly attributed to
the red wing of relativistic lines.

The Seyfert-1 galaxy NGC 3516 is an excellent case study.  {\it ASCA}
spectra revealed a notable relativistic line with a prominent red wing
(see, e.g., Nandra et al.\ 1997; Reynolds 1997).  An initial analysis
of observations with {\it Chandra} and {\it XMM-Newton} confirmed the
disk line, and also revealed evidence for variable narrow emission
lines that may be related to orbital motions (Turner et al.\ 2002).
In a later analysis of the {\it Chandra}/HETGS spectrum, it was found
that a warm/hot absorber capable of creating curvature in the Fe~K
band was statistically permissible (Turner et al.\ 2005).  In fact, if
one looks closely at Figure 6 in Turner et al.\ (2005), it is clear
that this model greatly over-predicts the observed absorption in the
Fe~K band (see Figure 12).  The absorption model does not fit the
data, strongly suggesting the observed curvature above 3~keV in NGC
3516 is actually due to the red wing of a relativistic line.  The
broad-band continuum is defined better in {\it Suzaku} observations of
NGC 3516, leaving less uncertaingy in the low energy spectrum.  A
relativistic disk line is required in the initial {\it Suzaku}
spectrum at the 99.999\% level of confidence, even when low-energy
absorption from a warm absorber is included (Markowitz et al.\ 2006).

A careful analysis of a 522~ksec {\it Chandra}/HETGS spectrum of
MCG-6-30-15 also rules out absorption creating or affecting the red
wing; such absorption models predict a strong line in the Fe~K band
that is not observed (Young et al.\ 2005).  Please see Figure 13 to
see how absorption fails to fit the observed spectrum.  This careful
study follows from a thorough analysis of the 320~ksec {\it
XMM-Newton} spectrum of MCG-6-30-15.  Vaughan \& Fabian (2004)
conducted an extremely rigorous and broad-ranging analysis which
examined many possible ways a line-like feature might arise (including
absorption), and conclude that the relativistic line in MCG-6-30-15 is
robust.  This paper is recommended to the reader as perhaps the most
complete paper on the topic, and an excellent example of a rigorous
analysis.

The most extreme example of complex low-energy absorption in a
Seyfert-1 galaxy is likely NGC 3783.  In an analysis of a 900~ksec
{\it Chandra}/HETGS spectrum, Kaspi et al.\ (2002) report the
detection of a Compton edge in the narrow component of the Fe~K line,
but do not consider a global spectral model appropriate for broad line
studies.  In an analysis of a deep {\it XMM-Newton} spectrum, Reeves
et al.\ (2004) find that absorption may affect the Fe~K range; this
agrees with an analysis of the long {\it Chandra} exposure by Yaqoob
et al.\ (2005) which only finds strong evidence for a small red extent
to the Fe K line.

Even in the extreme case of NGC~3783, however, when absorption and
relativistic disk line components are both included in spectral models
and the data alone decide the relative importance of these features, a
relativistic disk line is strongly required (Nandra et al.\ 2006).
Indeed, as noted above, the recent and {\it systematic} effort to fit
only the absorption strongly required by individual spectra finds that
relativistic lines are required in the {\it XMM-Newton} spectra of NGC
2992, MCG-5-23-16, NGC 3516, NGC 3783, NGC 4051, NGC 4151, Markarian
766, and MCG-6-30-15 (Nandra et al.\ 2006).  This demonstrates that
even on a narrow energy band, a rigorous fitting procedure clearly
reveals relativistic lines.  The veracity of disk lines is even more
strongly demonstrated in recent {\it Suzaku} observations, which
clearly reveal relativistic lines above and beyond any absorption
concerns in NGC 4051, NGC 2110, and 3C 120 (Reeves et al.\ 2006), NGC
2992 (Reeves et al.\ 2007, Yaqoob et al.\ 2007), MCG-5-23-16 (Reeves
et al.\ 2006, Reeves et al.\ 2007), NGC 3516 (Reeves et al.\ 2007,
Markowitz et al.\ 2007), and MCG-6-30-15 (Reeves et al.\ 2006,
Miniutti et al.\ 2007).

Somewhat separate from the issue of modeling time-averaged spectra, it
has recently been found that relativistic iron lines are more
prominent in low flux states than in high flux states (see, e.g.,
Vaughan \& Fabian 2004).  This behavior is superficially consistent
with variable absorption affecting the spectrum.  If the apparent
prominence of disk lines at low flux is actually caused by absorption,
however, variability should be seen in discrete absorption features.
In an analysis of the absorption lines in the 320~ksec {\it
XMM-Newton} spectrum of MCG-6-30-15, A. K. Turner et al.\ (2004) find that
only 4 of 51 absorption lines are variable.  This dramatic {\it lack}
of absorption variability strongly suggests that the observed spectral
curvature is due to a relativistic disk line and broad-band disk
reflection spectrum.

% ......................................................................

\subsection{Scattering Effects and Broad Emission Lines}

It has sometimes been argued that the breadth of Fe~K lines is not caused by
a combination of Doppler shifts and gravitational red-shifts in the
inner disk, but due to Comptonization (e.g., Misra \& Kembhavi 1998,
Misra \& Sutaria 1999).  It has also been suggested that the red wing
of Fe~K lines might be due to relativistic ($v/c \simeq 0.3$),
optically-thick, wide-angle or quasi-spherical outflows (see, e.g.,
Titarchuk, Kazanas, \& Becker 2003; Laming \& Titarchuk 2004; Laurent
\& Titarchuk 2007).  In both cases, a preponderance of observational
evidence and sound physical arguments rule-out these possibilities,
and demand that broad lines are shaped primarily by dynamics.

The case of Compton broadening of the line in MCG-6-30-15 as described
in the models of Misra \& Kembhavi (1998) and Misra \& Sutaria (1999)
is treated in detail in Reynolds \& Wilms (2000) and Ruszkowski et
al.\ (2000).  To create a red wing, line photons must be primarily
down-scattered, necessitating a Comptonization region around the
central engine with $\tau = 4$ and $kT \leq 0.5$~keV.  Several sound
arguments rule-out this possibility: (a) the spectral break expected
at $E_{br} = m_{e} c^{2} / \tau^{2} \simeq$30--40~keV (Guainazzi et
al.\ 1999) is not observed; (b) variability is observed corresponding
to length scales too short to be observed given the optical depth; and
(c) the (unobserved) blackbody component required to sustain the
Compton region would violate the blackbody limit for reasonable size
scales of the Comptonization region.  Turner et al.\ (2002) have also
pointed out that the detection of narrow features within the broad
line in NGC~3516 argues against strong Comptonization.  The same
argument can be extended to any source with evidence for variable
narrow emission lines, such as Mrk 766 (Turner et al.\ 2006).

The possibility of scattering creating red wings as per the models of
Titarchuk, Kazanas, \& Becker (2003), Laming \& Titarchuk (2004), and
Laurent \& Titarchuk (2007) is discussed briefly in Miller et al.\
(2004a), but numerous other studies also bear directly on such models.
Many observations clearly reveal X-ray flux variability on timescales
corresponding to the innermost stable circular orbit around black
holes, which is inconsistent with this region being screened by an
optically--thick outflow.  The predictions of simple reflection models
are strongly confirmed in low continuum flux phases, and where the
line flux does not trace the continuum flux, observations show that
absorption and scattering cannot be the explanation.  For examples,
please see, e.g., Gilfanov, Churazov, \& Revnivtsev 2000; Miniutti \&
Fabian 2004; Reynolds et al.\ 2004; Ponti et al.\ 2004; Vaughan \&
Fabian 2004; Iwasawa, Miniutti, \& Fabian 2004; Miller et al.\ 2004a;
and Miller \& Homan 2005.  Moreover, to achieve the degree of
scattering required to produce the red wings observed in Fe~K line
spectra, highly super-Eddington outflows (as much as 10 times the
Eddington rate) are required.  The outflows implied in stellar mass
black holes such as GX~339$-$4 are generally a small fraction of the
Eddington mass accretion rate (Miller et al.\ 2004b, 2006b), and those
seen in all black hole spectra fall well-below the velocity required
($v/c \geq 0.3$) for such effects to be important.
 
%----------------------------------------------------------------------

\section{An Informal Census of Relativistic Disk Lines}

The aim of this section is to collect an up-to-date census of sources
with relativistic disk line detections.  Evidence of C, N, and O disk
lines is only reported in sources that have compelling Fe disk lines,
so it is only necessary to count Fe lines.  Reported lines with
FWHM$\leq 0.1c$ are not considered, in order to ensure disk-like
velocities.  Similarly, detections of single {\it narrow} red or
blue-shifted lines are not considered as they may be related to jet
ejection events rather than disks.  All of the stellar-mass black hole
sources in the Remillard \& McClintock (2006) catalog are considered;
black hole ``candidates'' are not separated from
dynamically-constrained systems because the phenomena observed in
``candidate'' sources strongly suggests black holes.  The survey
undertaken by Nandra et al.\ (2006) considered only Seyfert AGN with
more than 100,000 X-ray photons detected with {\it XMM-Newton}, and
made new spectral fits.  The exposure threshold and uniformity of the
Nandra et al.\ (2006) survey ensure that its estimate of the fraction of
AGN with a broad disk line is robust.  In contrast, the census undertaken
here does not place a limit on how well a source has been observed,
and relies on fits reported in the literature.  However, the list
generated here is broad and may be a useful starting point for deeper
observations and new studies.

As discussed at length above, the best examples of relativistic disk
lines represent extremely incisive probes of black hole spin and the
strong gravitational environment close to a black hole.  The power of
the the best detections is greatest, though, when we remain critical
of all detections.  In this spirit, reported disk lines have been
divided into three groups: Tier 1, Tier 2, and Tier 3, from the most
robust and incisive lines to those detected with less certainty.
Because it is practical to do so, significant non-detections in
stellar-mass black holes are also listed.  The standards used to
distinguish these groups cannot be exclusively quantitative because
different scientists have employed different models, different
instruments, and different statistical methodology.  Moreover, the
standards applied to lines from disks around stellar-mass black holes
and disks in Seyfert AGN differ because of their intrinsically
different low energy spectra, fluxes, and timescales.  The criteria
below describe the lines in each tier:\\
 
\noindent {\it Seyfert AGN, Tier 1}: Strong lines revealed in deep
{\it Chandra} and/or {\it XMM-Newton} and/or {\it Suzaku} observations
that are robust against plausible absorption and reveal possible
evidence of black hole spin.

\noindent {\it Seyfert AGN, Tier 2}: Lines seen with {\it XMM-Newton},
{\it Suzaku}, or with {\it BeppoSAX} in shorter observations than
sources in Tier 1.  Line asymmetry may be less clear than Tier 1.

\noindent {\it Seyfert AGN, Tier 3}: Weak line detections, whether due
to intrinsic weakness or very short observations.\\

\noindent {\it Stellar-mass black holes, Tier 1}: Lines detected with
multiple instruments (if not an instrument with at least CCD spectral
resolution, then an imaging mission with low energy coverage and a
mission with broad energy range) and in multiple states with a clear
asymmetric profile.

\noindent {\it Stellar-mass black holes, Tier 2}: Lines detected with
multiple instruments, and/or in multiple states, and with an imaging
mission with low energy coverage.

\noindent {\it Stellar-mass black holes, Tier 3}: Lines detected with
{\it RXTE} only.  Sources observed only with {\it Ginga} are not
included.

\noindent {\it Significant non-detections}: Sources
observed with a recent soft X-ray spectrometer covering the Fe~K band,
observed in states with strong hard flux, fit with
now-standard continuum models, and in which observations can
rule-out broad lines with an equivalent width of 180~eV (as per
unity disk covering factor, George \& Fabian 1991).\\

Analysis by Nandra et al.\ (1997) found disk lines in 77\% of
Seyfert-1 AGN observed with {\it ASCA}.  Moreover, this work reported
a strong disk line profile in the average spectrum of Seyfert AGN after
excluding clear individual detections, further suggesting that such
lines are common.  The absence of strong detections in short initial
observations with {\it Chandra} and {\it XMM-Newton} generated some
concern about the veracity of lines detected with {\it ASCA}.  Thus,
it is worth looking at the Nandra et al.\ (1997) results critically.
Table 3 in Nandra et al.\ (1997) reports the results of fits to 23
spectra from 18 Seyfert AGN.  A single broad Gaussian was used to
model Fe disk lines.  But how many of these detections are robust?  A
modest requirement on the line width might be: $FWHM/err(FWHM) \geq
2$.  Only eight sources pass this cut.  This certainly does not mean
that only 8 of 18 sources can be expected to have disk lines, on
average.  However, it does mean that as of 1997, {\it ASCA}
observations had only yielded modestly convincing detections of
relativistic disk lines in eight sources.  The fact that the new
survey detailed in Nandra et al.\ (2006) finds relativistic disk lines
in 73\% of Seyferts -- though more conservative modeling is employed
and an exposure criterion is enforced -- strongly confirms that lines
are commonly found in deep observations.  This finding is supported by
Guainazzi, Bianchi, \& Dovciak (2006).

The results of the informal Seyfert census are given in Table 1.  The
number of strong disk line detections in Seyfert AGN is clearly {\it
growing} in the {\it Chandra}, {\it XMM-Newton}, and {\it Suzaku} era.
There are now nine sources with very strong disk line detections, a
further six sources with compelling evidence for disk lines, and 15
sources with detections that need to be confirmed and investigated
more deeply.  If we take each of these detections at face value, then
the 18 sources reported by Nandra et al.\ (1997) has increased to
30.  Many prior detections with {\it ASCA} are confirmed and some new
detections have been made.  In a small number of cases, lines detected
with {\it ASCA} are no longer detected.  Given that line variability
patterns prove to be complex, it is likely that some current
non-detections are due to intrinsic variability.   In some cases, prior
detections with {\it ASCA} may have been erroneous.

The results of the informal census of stellar-mass black holes are
given in Table 2.  The ability to obtain moderate and high resolution
spectra has transformed the field, bolstered prior claims based on gas
detector spectra, and inspired new archival searches.  Prior to {\it
Chandra} and {\it XMM-Newton}, the line profile in Cygnus X-1 was
likely the only case widely regarded as robust.  In the current era,
there are 16 sources in which compelling lines have been detected, of
19 in which lines could have been detected.  Of eight sources observed
well at CCD or gratings resolution, six show relativistic disk line
profiles.  Thus, relativistic lines are detected in 75--85\% of
stellar-mass black holes, in strong agreement with the detection
fraction in Seyfert-1 AGN (Nandra et al.\ 1997; Nandra et al.\ 2006).

It is difficult to detect continuum emission or lines from the inner
disk in Seyfert-2 AGN, due to obscuration from the torus.  It is also
difficult to detect relativistic disk lines in stellar-mass black
holes that are viewed at high inclinations.  While {\it Chandra}
observations of XTE~J1118$+$480 did not achieve the sensitivity
required to detect or reject disk lines, sensitive observations have
been made of other edge-on black holes.  A disk line has never been
convincingly detected in 4U 1630$-$472, and simultaneous {\it Chandra}
and {\it RXTE} observations of H 1743$-$322 did not reveal disk line
emission with tight limits (Miller et al.\ 2006b).  There is likely no
geometry in stellar-mass black holes that accurately mimics the torus,
but obscuration from the outer accretion disk and/or enhanced
scattering effects due to the edge-on viewing angle may act to limit
our ability to detect disk lines.

%----------------------------------------------------------------------

\section{Constraints on the Innermost Accretion Flow Geometry}

The nature of hard X-ray emission in black holes is an outstanding
problem.  At present, it is not clear which process gives the dominant
contribution to the observed X-ray emission.  The geometry of the
emitting region(s) is an closely related problem.  A central
Comptonizing ``corona'', magnetic flaring loops above the disk, and
processes in the base of a jet may all play a role.  Indeed, these
possibilities may not even be distinct -- they may all be related and
part of a larger process.  Relativistic disk lines and the larger disk
reflection spectrum arise through the interaction of the hard X-ray
region and the inner disk.  As a result, these features can both serve
to measure the inner disk radius and to constrain the nature of the
hard X-ray emission region.

Recent disk line studies provide growing evidence that hard X-ray
emission may be incredibly central and compact (somewhat arbitrarily,
within $10~r_{g}$ or less), at least at reasonably high mass accretion
rates (e.g., $L_{X}/L_{Edd.} \geq 0.01$).  Steep line emissivity
profiles have now been measured in a number of Seyfert AGN and
stellar-mass black holes (e.g., Wilms et al.\ 2001, Miller et al.\
2002b), indicating strongly centralized energy dissipation.  While
this is consistent with a torque at the inner disk (perhaps through
magnetic connections to the ergosphere or matter in the plunging
region), the success of light bending models for X-ray variability
suggests that a small source only a few $r_{g}$ above the hole may
provide much of the hard X-ray emission (e.g., Miniutti \& Fabian
2004).  In turn, this suggests that only a few flares (or perhaps the
base of a jet) may dominate hard X-ray production in actively
accreting black holes (Uttley \& McHardy 2001, Uttley, McHardy, \&
Vaughan 2005; for a treatment of jets and coronae, see Markoff, Nowak,
\& Wilms 2005).

The variability seen in narrower components in the line profiles
observed in NGC 3516 and Mrk~766 (Iwasawa, Miniutti, \& Fabian 2004;
Turner et al.\ 2006) is also consistent with relatively small magnetic
flares.  These observations suggest that flares need not be extremely
centralized, and that flares can co-rotate above the disk for the
duration of an orbit (or more).  It is not clear if the line
variability seen in these systems is plausibly consistent with
variable reflection due to a warp, as may be the case in
GRS~1915$+$105 (Miller \& Homan 2005).

The nature of the innermost accretion flow at low accretion rates is
also a matter of some uncertainty.  Some models suggest that the
innermost disk may radially recede, and be replaced by an
advection-dominated inner flow.  Recent analysis of relativistic disk
line and continuum spectra from the stellar-mass black holes Cygnus
X-1 and GX~339$-$4, and the black hole candidate SWIFT J1753.5$-$0127
in the ``low/hard'' state suggests that disks instead remain at or
close to the ISCO for $L_{X}/L_{Edd} \geq 0.003$ (Miller et al.\
2006b).  Low-luminosity AGN like M81 and NGC 4258 offer the chance to
study inner flow geometries at much lower fractions (roughly 100--1000
times lower) of the Eddington luminosity.  The possibly relativistic
line profile in M81 appears to be a collection of narrow lines,
perhaps indicating that a thin disk does not extend to the ISCO
(Dewangan et al.\ 2004).  At present, differing results from analyses
of the Fe line in NGC 4258 do not give a coherent picture of this
source (Reynolds, Nowak, \& Maloney 2000; Fruscione et al.\ 2005).

% ......................................................................

\section{Summary and Future Prospects}

Robust explorations and tests of exotic scientific predictions require
well--understood tools that will serve to generate as many trials as
possible.  Whether exploring General Relativity, or how black holes and
galaxies co-evolve, relativistic disk lines satisfy this criteria.  
Observations being made with current observatories show that the next
generation of X-ray missions are poised to fulfill their enormous
promise.

In recent years, the utility of relativistic disk lines has been
tested and confirmed.  Current observations clearly show that deep
observations require relativistic disk lines in a high fraction of
Seyfert-1 AGN, and the number of sources in which disk lines have been
detected continues to grow.  The fact that disk lines are observed in
both supermassive and stellar-mass black holes facilitates many
fundamental comparisons.  Moreover, recent theoretical advances have
kept pace: the recent development of variable-spin line models is
ideally timed to coincide with the launch of {\it Suzaku}, and new
deep {\it XMM-Newton} observations.

Thus, although spin constraints were possible in the past, realistic spin
measurements are possible now, and it is essential that the field
pursue such measurements in the next few years.  Although many line
profiles and variability trends strongly suggest high black hole spin
parameters in a number of sources, this is only confirmed in the case
of MCG-6-30-15 (Brenneman \& Reynolds 2006).  Deep {\it XMM-Newton}
and {\it Suzaku} observations and systematic measurements of spin
parameters are urgently needed.  Long observations of Tier 2 and Tier
3 Seyferts, and of all new transient stellar-mass black holes, are
required to build statistics.

There are already important indications that future missions like {\it
Constellation-X} and {\it XEUS} will be able to measure spin in a
large number of black holes, and to probe the spin history of AGN out
to moderate red-shift.  Streyblyanska et al.\ (2005) examined deep
{\it XMM-Newton} observations of the Lockman Hole; after adding the
spectra of AGN in similar red-shift bins out to $z=2$, relativistic
disk lines are revealed in both the average Seyfert-1 and Seyfert-2
spectra (please see Figure 14).  A separate analysis of {\it
XMM-Newton} and {\it Chandra} deep fields reaches similar conclusions
(Comastri, Brusa, \& Gilli 2006).  Current predictions suggest that
70\% of all supermassive black holes should have maximal spin
(Volonteri et al.\ 2005), and only missions such as {\it
Constellation-X} and {\it XEUS} will be able to test such models and
better our understanding of how black holes and galaxies co-evolve.

Quite apart from time-averaged measurements of spin, iron line
``reverberation mapping'' offers an excellent opportunity to probe
strong-field gravitational effects (Reynolds et al.\ 1999, Young \&
Reynolds 2000).  With sufficiently high collecting area, it becomes
possible to watch the disk respond to individual hard X-ray flares
through the evolution of iron line emission.  The flux pattern
observed (the ``transfer function'') encodes precise information on the
black hole spin and the nature of inner accretion flow.  New results
suggesting orbiting flares (e.g., Iwasawa, Miniutti, \& Fabian 2004),
gravitational light bending (e.g., Miniutti \& Fabian 2004), and
centrally-concentrated X-ray emission all bode extremely
well for reverberation mapping.  For a broad range of configurations,
{\it Constellation-X} and {\it XEUS} will be able to perform
reverberation mapping in approximately 10 Seyfert AGN in addition to
the hundreds of spin determinations possible via fits to time-averaged
line profiles.

%------------------------------------------------------------------------

\section*{Acknowledgments}

I would like to thank all of my colleagues for their hard work in this
area.  I would especially like to acknowledge the following people for
their comments and generosity: Andy Fabian, Jeroen Homan, Kazushi
Iwasawa, Julian Krolik, Alex Markowitz, Giovanni Miniutti, Paul
Nandra, Michal Nowak, James Reeves, Chris Reynolds, Jane Turner, and
Simon Vaughan.  Everyone active in this field is indebted to Fred
Jansen, Kazuhisa Mitsuda, Arvind Parmar, Norbert Schartel, Jean Swank,
Harvey Tananbaum, and Nick White for directing missions (and planning
future missions) that make this work possible.

%------------------------------------------------------------------------

\begin{figure}
\centerline{\psfig{figure=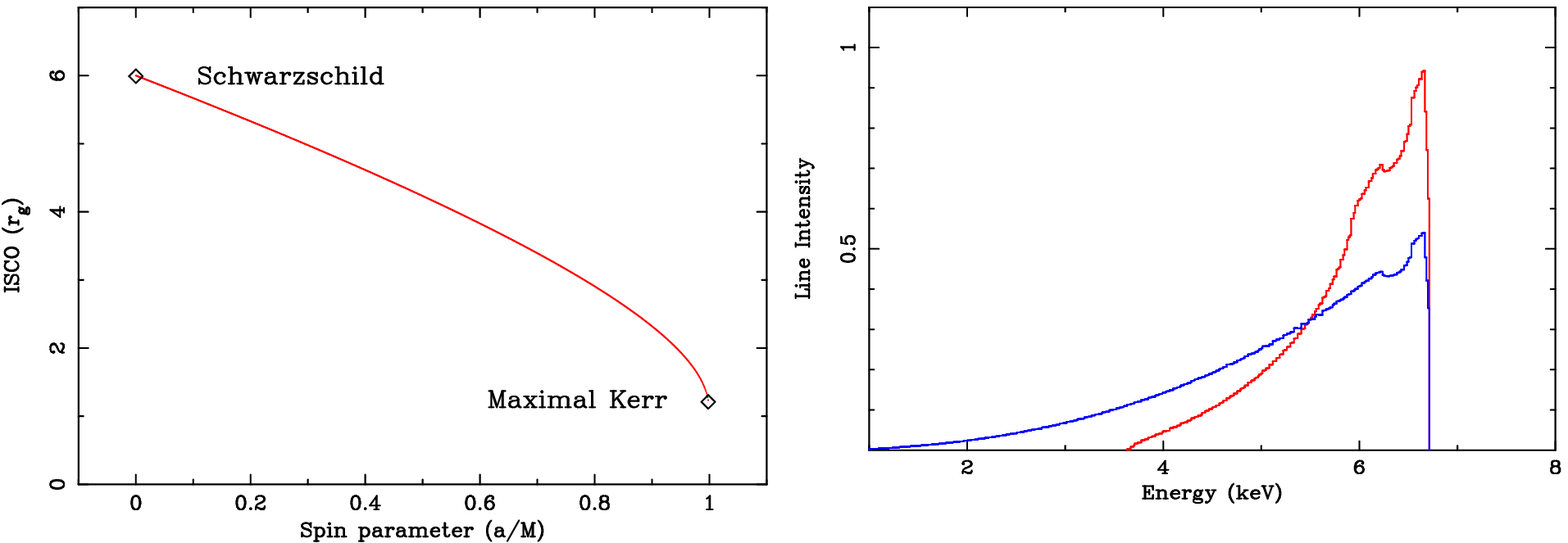,height=1.8in}}
\caption{{\it Left}: The dependence of the innermost stable circular
orbit (ISCO) on the black hole spin parameter is shown here, from
Schwarzschild ($a = 0$) to maximal Kerr ($a = 0.998$) solutions.  {\it
Right}: The line profiles predicted in the case of Schwarzschild (red)
and maximal Kerr (blue) black holes are shown here.  It is the extent
of the red wing and its importance relative to the blue wing that
allow black hole spin to be determined with disk lines.  (Adapted from
Fabian \& Miniutti 2006).}
\label{fig01}
\end{figure}

%----------------------------------------------------------------------

\begin{figure}
\centerline{\psfig{figure=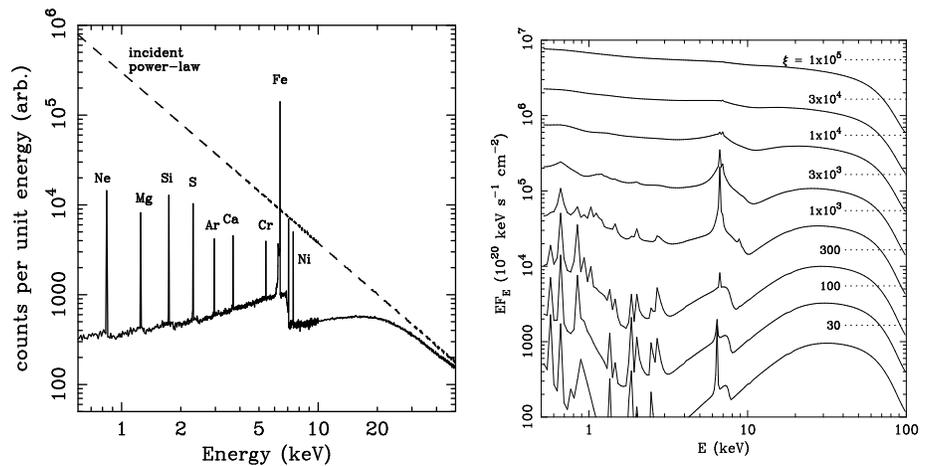,height=2.7in}}
\caption{{\it Left}: The response of a neutral slab of gas to an
incident power-law is shown here; this response spectrum is the disk
reflection spectrum.  (Adapted from Reynolds 1997).  {\it Right}: The
ionization of an accretion disk affects the reflection spectrum in
predictable ways.  Higher ionization results in greater broadening via
scattering in the disk atmosphere. (Adapted from Ross, Fabian, \& Young
1999.)}
\label{fig01}
\end{figure}

%----------------------------------------------------------------------

\begin{figure}
\centerline{\psfig{figure=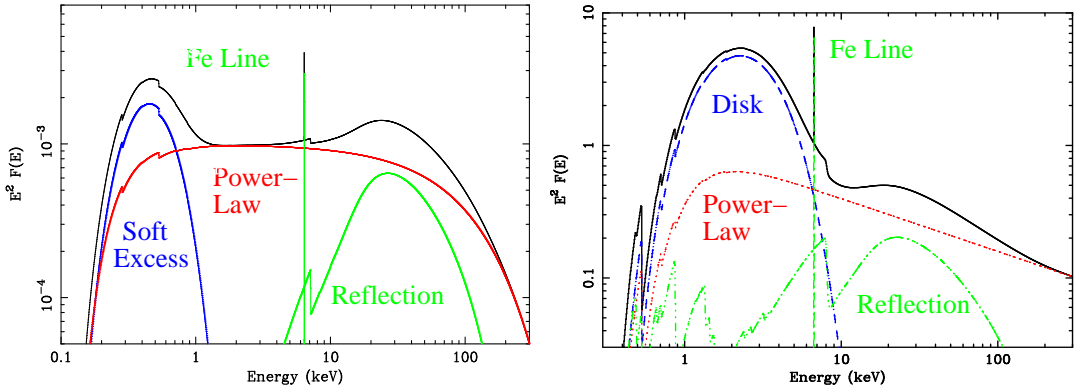,height=2.0in}}
\caption{Representative continuum and disk response spectra for a
Seyfert-1 AGN (left) and stellar-mass black hole (right) are shown
above.  The relativistic effects that shape the Fe line and reflection
have not included, so the Fe line is narrow.  In both cases, a
power-law is chosen to model the hard X-ray emission, but the
stellar-mass black hole power-law has a higher cut-off.  The effect of
neutral line-of-sight absorption in the Milky Way is also included in
both cases.  The ``soft excess'' component in the Seyfert-1 spectrum
is shown here as a blackbody but may arise through disk reflection;
please see the text for details.  (The left panel is adapted from
Fabian \& Miniutti 2006.)}
\label{fig01}
\end{figure}

%----------------------------------------------------------------------

\begin{figure}
\centerline{\psfig{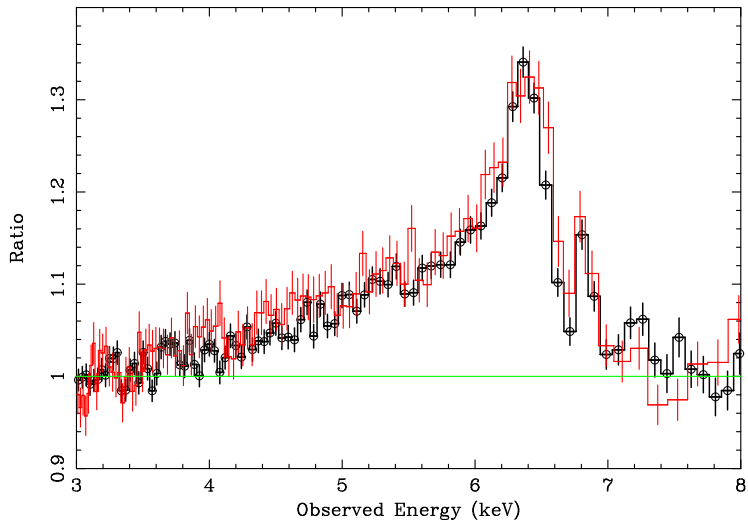}}
\caption{The figure above shows the relativistic disk line profile
revealed in MCG-6-30-15 after fitting for the continuum. (Adapted from
Miniutti et al.\ 2007 and Reeves et al.\ 2006.)  The line in
MCG-6-30-15 is the best example known presently, and these spectra above
are the best yet obtained.  The spectrum in black was obtained with
{\it Suzaku}, and the spectrum in red was obtained with {\it
XMM-Newton}.}
\label{fig01}
\end{figure}

%----------------------------------------------------------------------

\begin{figure}
\centerline{\psfig{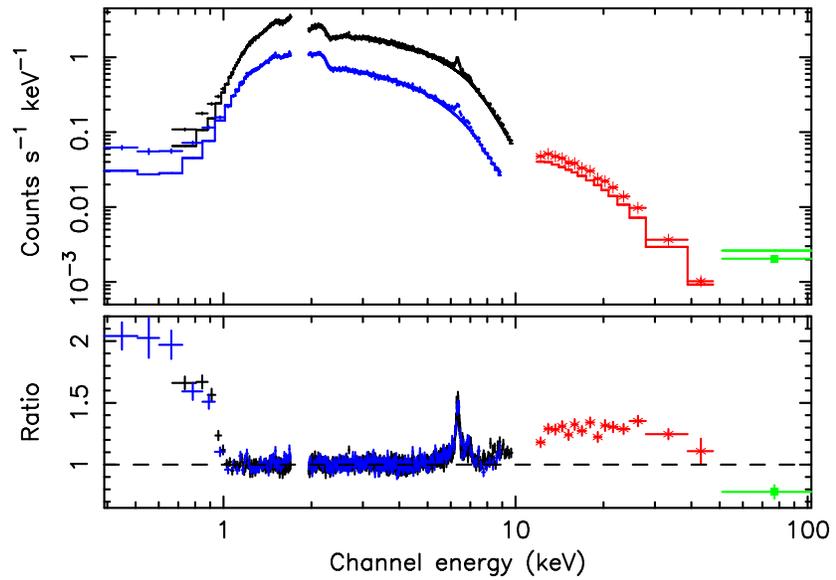}}
\caption{The figure above shows a {\it Suzaku} spectrum of the
Seyfert-1 AGN MCG-5-23-16 fit with a simple absorbed power-law (Reeves
et al.\ 2007).  The ratio of the data to this model is shown in the lower panel.
This figure demonstrates the ability of {\it Suzaku} to reveal all of
the key features of Seyfert X-ray spectra simultaneously, including
the soft excess, narrow iron line from distant reflection, and broad
iron line and reflection from the inner disk.  (The different colors
merely reflect spectra obtained using different cameras.  The figure
was adapted from Reeves et al.\ 2007.)}
\label{fig01}
\end{figure}

%----------------------------------------------------------------------

\begin{figure}
\centerline{\psfig{figure=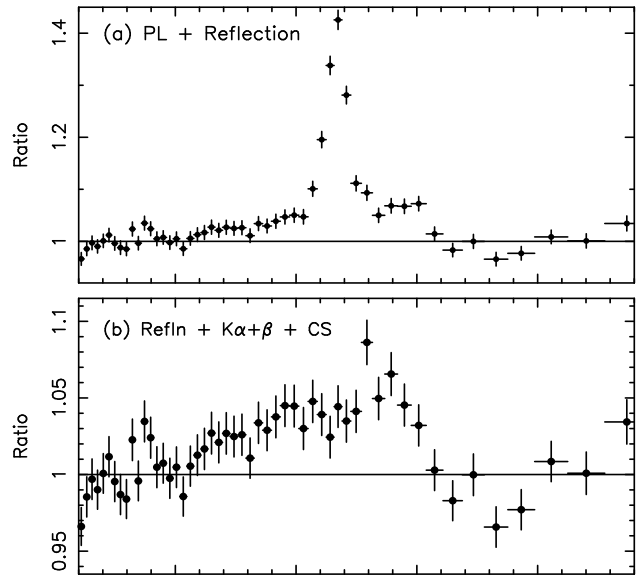,height=3.0in}}
\caption{The figure above shows the data/model ratio resulting from
more sophisticated fits to the spectrum of MCG-5-23-16 (Reeves et al.\
2007).  The top panel reveals narrow and broad Fe~K lines after
fitting for the continuum and disk reflection.  The bottom panel shows
the relativistic disk line that is revealed after fitting for narrow
Fe~K$\alpha$ and Fe~K$\beta$ lines.  (The figures above are adapted
from Reeves et al.\ 2007.)}
\label{fig01}
\end{figure}

%----------------------------------------------------------------------

\begin{figure}
\centerline{\psfig{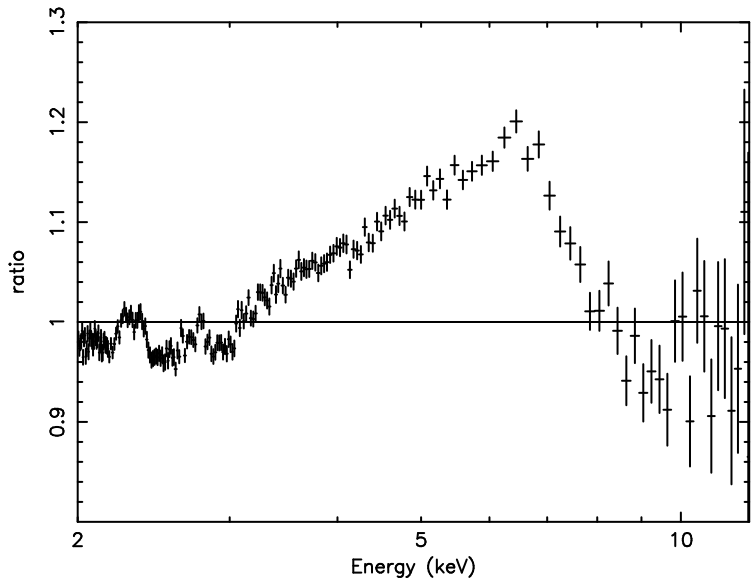}}
\caption{The relativistic Fe line revealed in an {\it XMM-Newton}
spectrum of the stellar-mass black hole GX 339$-$4 in a ``Very High''
state is shown above.  Fits with phenomenological and physical models
suggest that GX~339$-$4 harbors a black hole with $a \geq 0.9$.
(Adapted from Miller et al.\ 2004a).  The red wing of the line profile
seen above is extreme, and similar to that observed in MCG-6-30-15.}
\label{fig01}
\end{figure}

%----------------------------------------------------------------------

\begin{figure}
\centerline{\psfig{figure=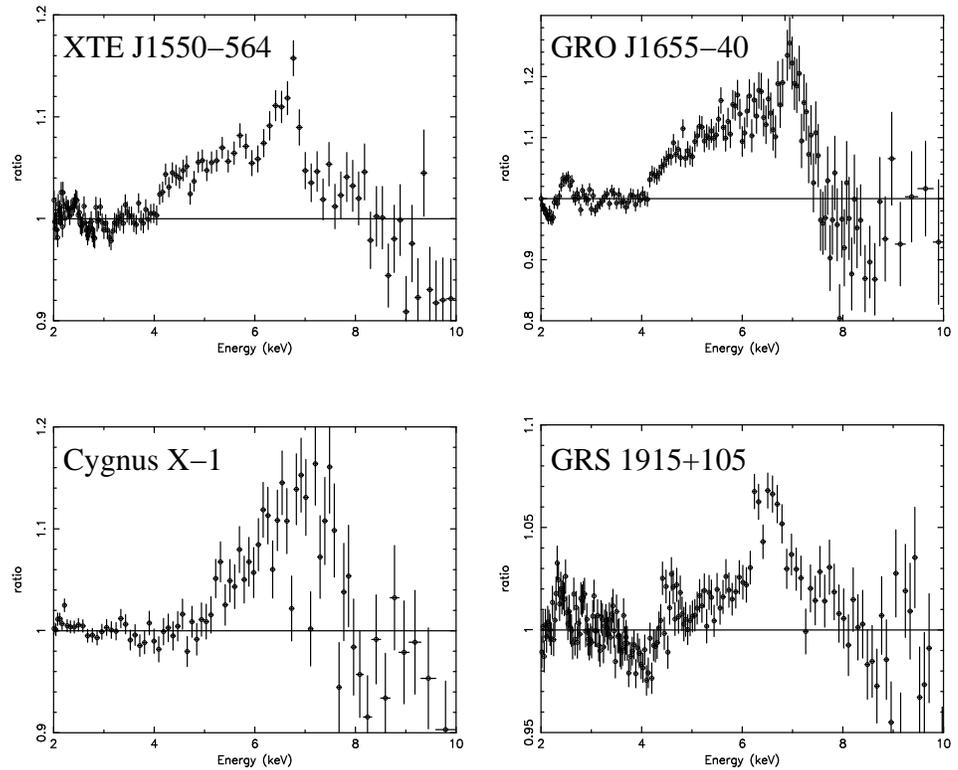,height=4.0in}}
\caption{The figure above shows the relativistic disk line profiles
revealed in an analysis of archival {\it ASCA}/GIS spectra of
stellar-mass black holes that were observed in bright phases.  (Adapted
from Miller et al.\ 2005).  The line profile in GRS~1915$+$105 is
clearly not as skewed as the others, and does not strongly require
black hole spin.}
\label{fig01}
\end{figure}

%----------------------------------------------------------------------

\begin{figure}
\centerline{\psfig{figure=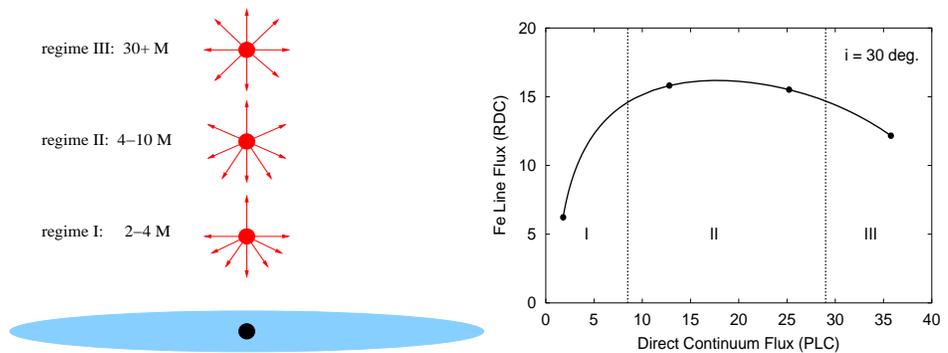,height=1.8in}}
\caption{{\it Left}: This panel depicts the ability of gravitational
light bending to focus emission from a constant-luminosity source at
different source heights (above, $r = GM/c^{2}$, with $G=c=1$, and the
figure is not shown to scale).  Close to the black hole, the source
emission is far from isotropic.  {\it Right}: Whereas a linear
relation between Fe line flux and ionizing continuum flux is expected
in the absence of light bending, a distinctly non-linear relation is
expected when light bending is important, as shown in the panel on the
right.  (Adapted from Miniutti \& Fabian 2004.)  The regimes defined in
this panel are depicted in the panel shown at left.}
\label{fig01}
\end{figure}

%----------------------------------------------------------------------

\begin{figure}
\centerline{\psfig{figure=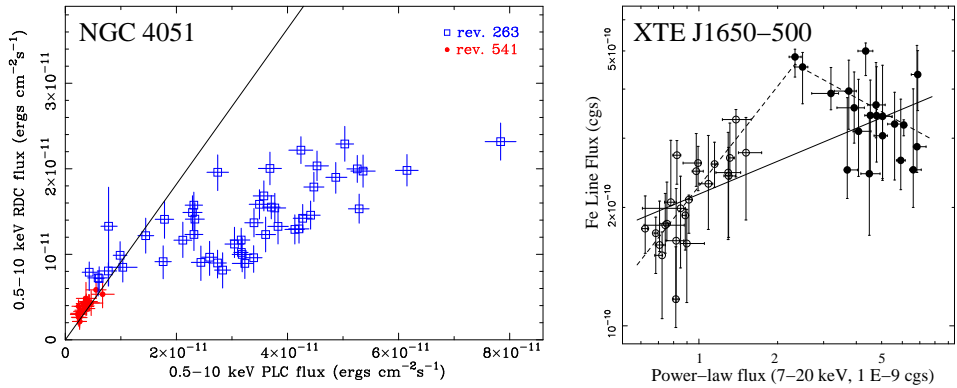,height=2.0in}}
\caption{{\it Left}: The relationship between the flux in the disk
reflection component and the irradiating power-law flux in the
Seyfert-1 AGN NGC 4051 is shown here.  (Adapted from Ponti et al.\
2006.)  {\it Right}: The relationship between Fe line flux and the
disk-irradiating power-law flux in the stellar-mass black hole XTE
J1650$-$500 is shown in this panel.  (Adapted from Rossi et al.\ 2005.)
In each case, it is clear that the line/reflection and power-law
fluxes are directly related at low fluxes.  The overall trends shown
here are consistent with the flux relations predicted by models for
gravitational light bending close to spinning black holes (e.g.,
Miniutti \& Fabian 2004).  A growing number of Seyfert AGN display
similar trends; more work is needed in the case of stellar-mass black
holes.}
\label{fig01}
\end{figure}

%----------------------------------------------------------------------

\begin{figure}
\centerline{\psfig{figure=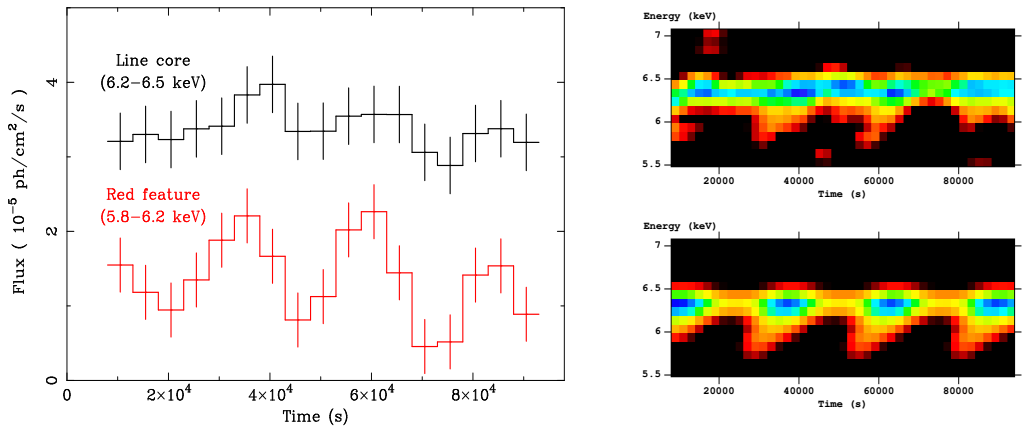,height=2.2in}}
\caption{{\it Left}: The lightcurves of components in the Fe~K band of
NGC 3516 is shown here.  One narrow line component in the Fe~K band is
especially variable.  {\it Above right}: In this figure, the
variations at left are shown as a trailed flux excess versus time map.
{\it Below right}: A theoretical model for the Fe line flux variations
based on a flare orbiting the black hole at radius of only 7--16
$r_{g}$ is shown here (Iwasawa, Miniutti, \& Fabian 2004; see also
Fabian \& Miniutti 2006).  The saw-tooth pattern is a direct result of
the extreme environment close to the black hole.  The remarkable
similarity of the data and model suggests that observations of
Seyferts with {\it XMM-Newton} and {\it Suzaku} can begin to probe
orbital timescale variability.  (The figures above were adapted from
Iwasawa, Miniutti, \& Fabian 2004.)}
\label{fig01}
\end{figure}

%----------------------------------------------------------------------

\begin{figure}
\centerline{\psfig{figure=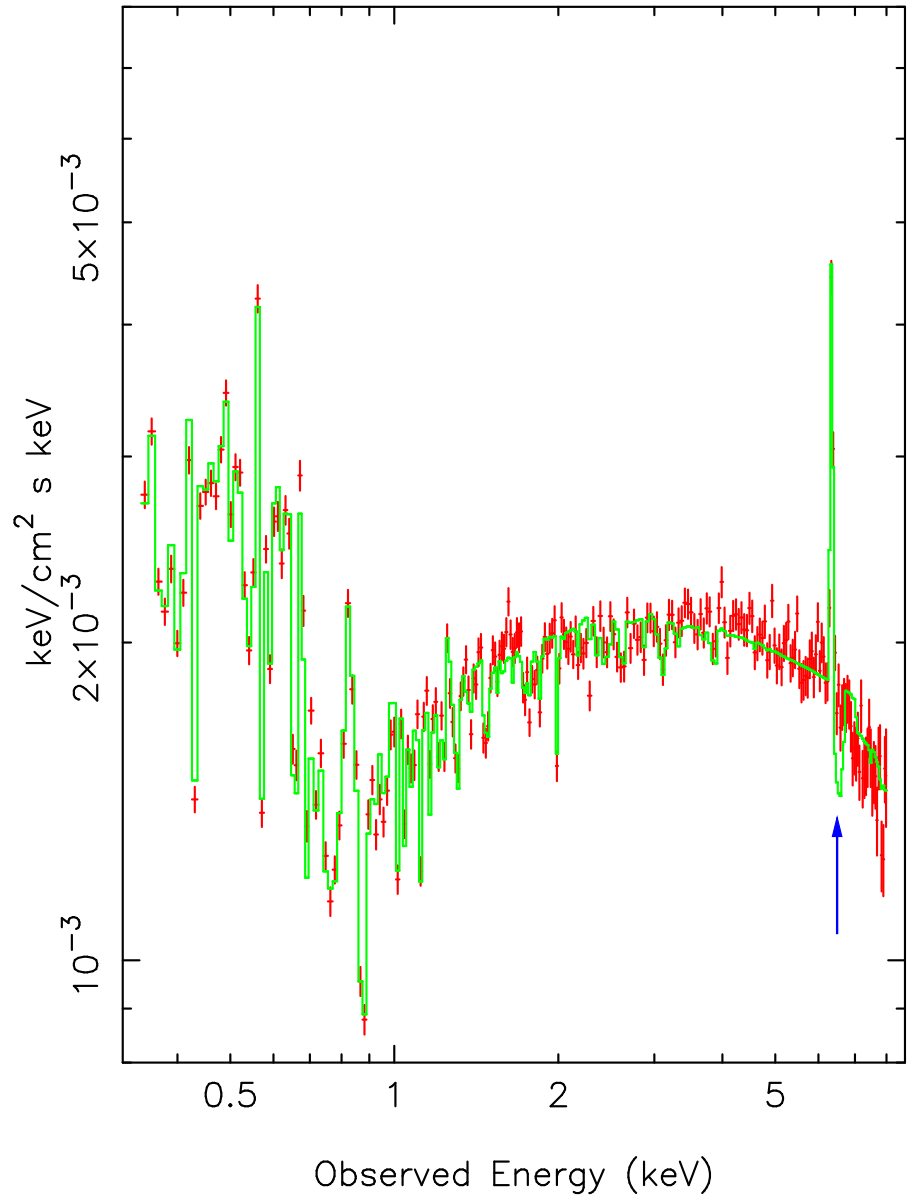,height=6.25in}}
\caption{An {\it XMM-Newton} spectrum of Seyfert-1 AGN NGC 3516 is
shown above, fit with a model including warm/hot absorption adding
curvature to the Fe~K band (Turner et al.\ 2004).  The blue arrow
emphasizes a low-ionization Fe absorption line predicted near to
6.4~keV that is not observed.  As with MCG-6-30-15, this spectrum
shows that relativistic lines, not absorption, must create the
curvature observed in the Fe~K band.  (The figure above was adapted
from Turner et al.\ 2004.)}
\label{fig01}
\end{figure}

%----------------------------------------------------------------------

\begin{figure}
\centerline{\psfig{figure=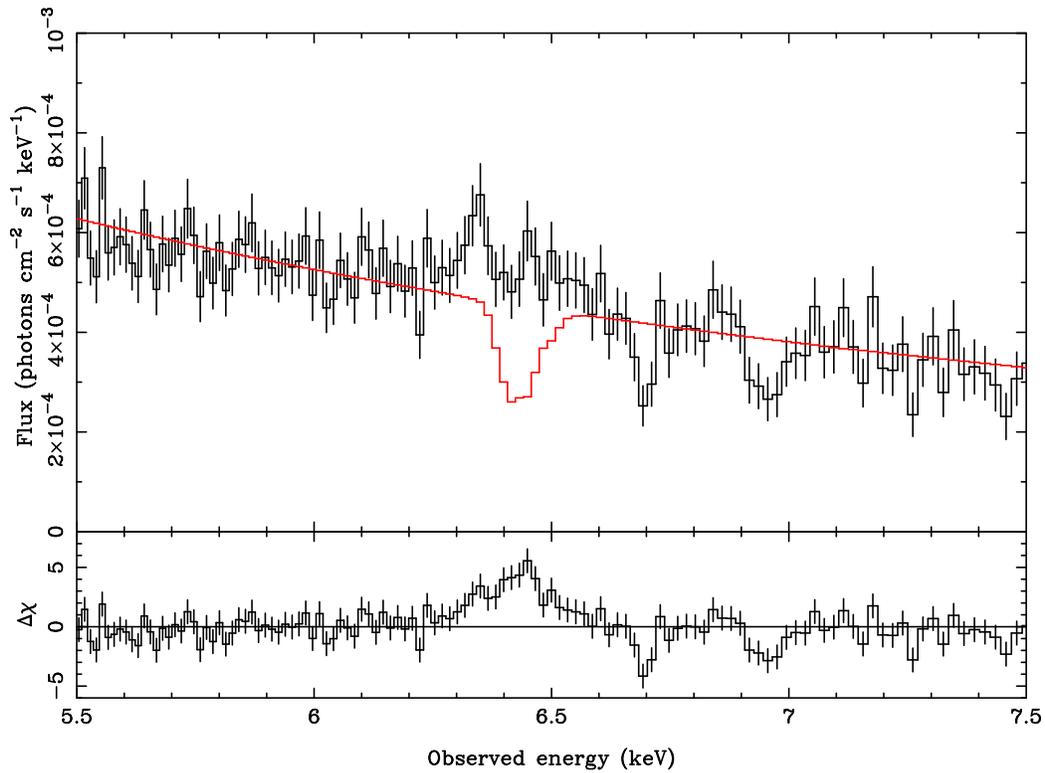,height=4.0in,angle=-90}}
\caption{The 522~ksec {\it Chandra}/HETGS spectrum of MCG-6-30-15 is
shown above, fit with a model that would explain curvature in the Fe K
band as absorption rather than a relativistic line.  The model
predicts a strong neutral absorption which is not observed, signaling
that curvature in the Fe K band is indeed due to relativistic disk lines
in Seyfert AGN (Young et al.\ 2005).  (The figure above was adapted
from Young et al.\ 2005.)}
\label{fig01}
\end{figure}

%----------------------------------------------------------------------
\begin{table}[t]

%\begin{footnotesize}
\begin{center}
\begin{tabular}{llll}
\multicolumn{4}{c}{Tier 1} \\
\hline
3C 120$^{1}$ & NGC 2992$^{2}$ & MCG-5-23-16$^{3}$ & NGC 3516$^{4}$ \\
NGC 3783$^{5}$ & NGC 4051$^{6}$ & NGC 4151$^{7}$ & Mrk 766$^{8}$ \\
MCG-6-30-15$^{9}$ & ~ & ~ & ~ \\
\hline
~ & ~ & ~ & ~ \\
\multicolumn{4}{c}{Tier 2} \\
\hline
Mrk 335$^{10}$ & Q 0056$-$363$^{11}$ & IRAS 06269$-$0543$^{12}$ & PG 1425$+$267$^{13}$ \\
Mrk 841$^{14}$ & IRAS 18325$-$5926$^{15}$ & ~ & ~ \\
\hline
~ & ~ & ~ & ~ \\
\multicolumn{4}{c}{Tier 3} \\
\hline
I Zw 1$^{16}$ & Fairall 9$^{17}$ & Mrk 359$^{18}$ & ESO 198-G24$^{19}$ \\
1H 0419-577$^{20}$ & Ark 120$^{21}$ & MCG-02-14-009$^{22}$ & 1H 0707-495$^{23}$ \\
PG 1211$+$143$^{24}$ & Mrk 205$^{25}$ & 3C 273$^{26}$ & 4U 1344-60$^{27}$ \\
IC 4329a$^{28}$ & NGC 5506$^{29}$ & II Zw 177$^{30}$  & ~ \\
\hline
\end{tabular}
\vspace*{\baselineskip}~\\ \end{center} 
\caption{Seyfert AGN in which relativistic disk lines have been
detected are listed above.  The sources are grouped into tiers
according to the nature of the detections; please see the text for
details.
$^{1}$ Reeves et al.\ (2006);
$^{2}$ Yaqoob et al.\ (2007), Reeves et al.\ (2006);
$^{3}$ Reeves et al.\ (2006,2007);
$^{4}$ Turner et al.\ (2002), Markowitz et al.\ (2006);
$^{5}$ Nandra et al.\ (2006);
$^{6}$ Nandra et al.\ (2006), Ponti et al.\ (2006), Reeves et al.\ (2006);
$^{7}$ Nandra et al.\ (2006);
$^{8}$ Mason et al.\ (2003), Nandra et al.\ (2006);
$^{9}$ Wilms et al.\ (2001), Fabian et al.\ (2002); Reynolds et al.\ (2004); Vaughan \& Fabian (2004); Miniutti et al.\ (2007);
$^{10}$ Gondoin et al.\ (2002), Longinotti et al.\ (2007);
$^{11}$ Porquet \& Reeves (2003);
$^{12}$ Gallo et al.\ (2007, in prep.);
$^{13}$ Miniutti \& Fabian (2006);
$^{14}$ Petrucci et al.\ (2002, 2006);
$^{15}$ Iwasawa et al.\ (2004);
$^{16}$ Gallo et al.\ (2004);
$^{17}$ Brenneman et al.\ (2007, in prep.);
$^{18}$ O'Brien et al.\ (2001);
$^{19}$ Porquet et al.\ (2004), simple independent analysis done for this review;
$^{20}$ Pounds et al.\ (2004), Fabian et al.\ (2005);
$^{21}$ Vaughan et al.\ (2004);
$^{22}$ Porquet (2006);
$^{23}$ Fabian et al.\ (2004);
$^{24}$ Pounds et al.\ (2003);
$^{25}$ Reeves et al.\ (2001);
$^{26}$ Turler et al.\ (2006);
$^{27}$ Piconcelli et al.\ (2006);
$^{28}$ Markowitz, Reeves, \& Braito (2006);
$^{29}$ Matt et al.\ (2001);
$^{30}$ Gallo (2006).}
\vspace{-1.0\baselineskip}
%\end{footnotesize}
\end{table}

%----------------------------------------------------------------------

\begin{table}[t]
%\begin{footnotesize}
\begin{center}
\begin{tabular}{llll}
\multicolumn{4}{c}{Tier 1} \\
\hline
XTE J1550-564$^{1}$ & XTE J1650$-$500$^{2}$ & GRO J1655$-$40$^{3}$ & GX 339$-$4$^{4}$ \\
SAX~J1711.6$-$3808$^{5}$ & GRS~1915$+$105$^{6}$ & Cygnus X-1$^{7}$ & ~ \\
\hline
~ & ~ & ~ & ~ \\
\multicolumn{4}{c}{Tier 2} \\
\hline
4U 1543$-$475$^{8}$ & V4641 Sgr$^{9}$ & 4U 1908$+$09$^{10}$ & XTE~J2012$+$38$^{11}$ \\
\hline
~ & ~ & ~ & ~ \\
\multicolumn{4}{c}{Tier 3} \\
\hline
GS 1354$-$64$^{12}$ & GRS 1737$-$31$^{13}$ & XTE~J1748$-$28$^{14}$ & XTE J1755$-$32$^{15}$ \\
XTE~J1859$+$226$^{16}$ & ~ & ~ & ~ \\
\hline
~ & ~ & ~ & ~ \\
\multicolumn{4}{c}{Significant Non-Detections} \\
\hline
4U 1630$-$472 & H~1743$-$322 & SW J1753.5$-$0127 & ~ \\ 
\hline
\end{tabular}
\vspace*{\baselineskip}~\\ \end{center} 
\caption{\small Stellar-mass black holes in which relativistic disk lines and sources with significant non-detections are listed above.  The sources are grouped into tiers according to the nature of the line detections; please see the text for details.
$^{1}$ Miller et al.\ (2003,2004), Sobczak et al.\ (2000);
$^{2}$ Miller et al.\ (2004), Miniutti, Fabian, \& Miller (2004);
$^{3}$ Miller et al.\ (2004), Diaz Trigo et al.\ (2006);
$^{4}$ Miller et al.\ (2004,2004,2006);
$^{5}$ Martocchia et al.\ (2002,2006), Miller et al.\ (2004);
$^{6}$ in 't Zand et al.\ (2002a), Sanchez-Fernandez et al.\ (2006);
$^{7}$ Miller et al.\ (2002,2006);
$^{8}$ Park et al.\ (2004), van der Woerd, White, \& Kahn (1989);
$^{9}$ Miller et al.\ (2002);
$^{10}$ in 't Zand et al.\ (2002b);
$^{11}$ Campana et al.\ (2002);
$^{12}$ Revnivtsev et al.\ (2000);
$^{13}$ Cui et al.\ (1997);
$^{14}$ Miller et al.\ (2001);
$^{15}$ Revnivtsev, Gilfanov, \& Churazov (1998);
$^{16}$ simple independent analysis done for this review;
$^{17}$ numerous sources;
$^{18}$ Miller et al.\ (2006);
$^{19}$ Miller, Homan, \& Miniutti (2006).}
%{\small Table 1: Stellar-mass black holes with relativistic disk lines
%detected in their X-ray spectra, grouped into tiers according to their
%properties (see the text).  The sources are listed in order of
%increasing right ascension from left to right, and then top to
%bottom.}
\vspace{-1.0\baselineskip}
%\end{footnotesize}
\end{table}

%----------------------------------------------------------------------

%----------------------------------------------------------------------

\begin{figure}
\centerline{\psfig{figure=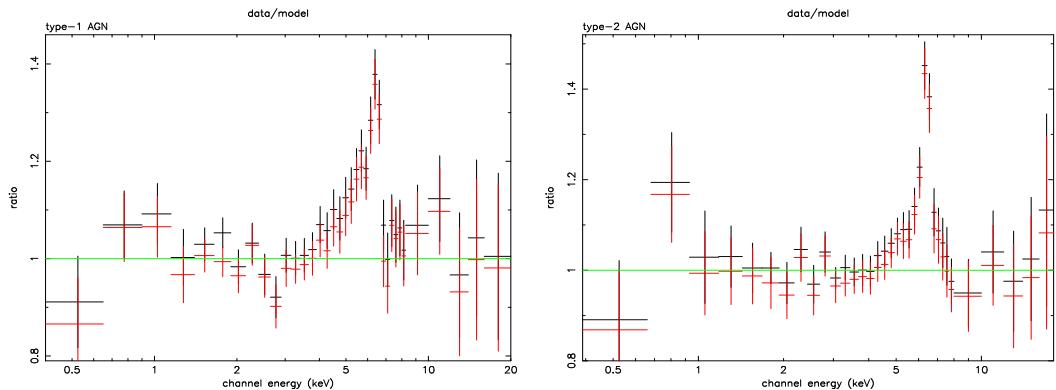,height=2.0in}}
\caption{Relativistic disk-line profiles are revealed in spectrum of
distant Seyfert-1 (left) and Seyfert-2 (right) AGN in a deep
observation of the Lockman Hole made with {\it XMM-Newton}
(Streblyanska et al.\ 2005).  Individual spectra from sources out to
$z \simeq 2$ were corrected to the rest frame and added to make the
spectra above.  This result makes it clear that future missions like
{\it Constellation-X} and {\it XEUS} will be able to study the spin
history of supermassive black holes.  (The different colors above
merely reflect spectra obtained with different cameras.  This figure
was adapted from Streblyanska et al.\ 2005.)}
\label{fig01}
\end{figure}

%----------------------------------------------------------------------

%%Unnumbered Literature Cited

% ----------------------------------------------------------------------

\newpage

\vspace*{-1.5 in}

%\centerline{\psfig{figure=aa43_brandt_table1.ps,height=11.0 in,angle=0}}

%------------------------------------------------------------------------

\end{document}